\documentclass[iop,apj,twocolappendix,revtex4]{emulateapj}
\usepackage{amstext}
\usepackage{graphicx}
\usepackage{pdflscape}
\usepackage{hyperref}

\shorttitle{The McGill Magnetar Catalog}
\shortauthors{Olausen \& Kaspi}

\begin{document}

\title{The McGill Magnetar Catalog}

\author{S. A. Olausen \& V. M. Kaspi}

\affil{Department of Physics, Rutherford Physics Building, McGill University,
3600 University Street, Montreal, Quebec, H3A 2T8, Canada}

\email{scott.olausen@mail.mcgill.ca}
\email{vkaspi@physics.mcgill.ca}

\begin{abstract}
We present a catalog of the 26 currently known magnetars and magnetar
candidates. We tabulate astrometric and timing data for all catalog
sources, as well as their observed radiative properties, particularly the
spectral parameters of the quiescent X-ray emission. We show histograms
of the spatial and timing properties of the magnetars, comparing them with
the known pulsar population, and we investigate and plot possible
correlations between their timing, X-ray, and multiwavelength properties.
We find the scale height of magnetars to be in the range 20--31\,pc, assuming
they are exponentially distributed. This range is smaller than that measured for
OB stars, providing evidence that magnetars are born from the most massive O
stars. From the same fits, we find that the Sun lies $\sim$13--22\,pc above the
Galactic plane, consistent with previous measurements. We confirm
previously identified correlations between quiescent X-ray luminosity
$L_{\mathrm{X}}$ and magnetic field $B$, as well as X-ray spectral
power-law index $\Gamma$ and $B$, and show evidence for an excluded
region in a plot of $L_{\mathrm{X}}$ vs.\ $\Gamma$. We also present an
updated $kT$ versus characteristic age plot, showing magnetars and
high-$B$ radio pulsars are hotter than lower-$B$ neutron stars of similar
age. Finally, we observe a striking difference between magnetars detected
in the the hard X-ray and radio bands; there is a clear correlation
between the hard and soft X-ray flux, whereas the radio-detected magnetars
all have low soft X-ray flux suggesting, if anything, that the two
bands are anti-correlated.

An online version of the catalog is located at \url{http://www.physics.mcgill.ca/~pulsar/magnetar/main.html}.
\end{abstract}


\section{Introduction}

The class of neutron stars today identified as ``magnetars'' was
first noted in 1979 with the detection of repeated bursts by space-based
hard X-ray/soft gamma-ray instruments \citep{mgi+79,mgg79,mg81}.
Though originally thought to have the same origin as the classical
gamma-ray bursts (GRBs), repeated bursts, including one enormous flare
on 1979 March 5, from the direction of the star-forming Dorado region
in the Large Magellanic Cloud \citep{mgi+79} as well as from what
today is known to be magnetar SGR 1900+14 \citep{mgg79,mg81}, provided
an important distinction and hint of a new class of Galactic sources.
The repeated bursts had somewhat softer spectra than those of most
GRBs, hence the sources' designation as `Soft Gamma Repeaters' (SGRs).
The 8-s pulsations seen in the declining flux tail following the large
flare were strongly suggestive of a neutron-star origin for SGRs.
That these two sources truly represented a distinct class of gamma-ray
bursters was more fully recognized in 1983 when a third Galactic source,
SGR 1806$-$20, underwent a major burst episode \citep{lfk+87}. Both
Galactic sources were noted to be very close to the Galactic Plane,
suggesting youth, a conclusion supported by the coincidence of the
LMC source with the supernova remnant N49 \citep{cdt+82}.

Meanwhile, \citet{fg81} reported an unusual 7-s X-ray pulsar, 1E
2259+586, in the Galactic supernova remnant CTB 109. Originally thought
to be a low-mass X-ray binary albeit without any obvious companion,
the source was soon recognized as being similar to a handful of other
`anomalous' sources (including 4U 0142+61 and 1E 1048.1$-$5937)
\citep[see][]{hel94, dt96, vtv95, ms95}, distinguished by their bright X-ray
pulsations at few-second periods, X-ray luminosities far greater than could
be explained via rotation power, but no apparent companions from which to
accrete. These distinctions led to the sources being termed
`Anomalous X-ray Pulsars' (AXPs) and this descriptor has stuck.

\citet{dt92} proposed that very strongly magnetized neutron stars
could be the origin of SGR emission, thereby coining the term `magnetar.'
\citet{td95} demonstrated that many SGR phenomena are readily explained
by a model in which spontaneous magnetic field decay serves as an
energy source for both the bursts and any persistent emission. They
cited not only energetics arguments but also the need for a high $B$
field to spin down a young neutron star from tens to hundreds of ms
(thought to be the typical birth spin period range) to several seconds,
within a supernova remnant lifetime. \citet{td96} further argued
that AXPs are also magnetars, with their X-ray luminosities powered
by magnetic field decay. The subsequent direct detection of spin-down
in an SGR at a rate consistent with the model prediction \citep{kds+98}
was a powerful confirmation of the magnetar picture. The detection
of SGR-like bursts from two AXPs \citep{gkw02,kgw+03} unified AXPs
and SGRs observationally, as predicted by \citet{td96}. Since then,
the distinction between AXPs and SGRs has been further blurred, with
practically all sources having shown characteristics of both: bursting
has now been shown to be a generic behavior of so-called AXPs \citep[e.g.][]{gkw04,wkg+05,kgk+10,sk11}
and AXP-like behavior (namely, absence of bursts for long periods)
has been seen in objects previously deemed SGRs, including the original
LMC SGR \citep{kkv+01}. It is clear that there exists a continuous
spectrum of behavior, ranging from anomalously high quiescent X-ray
luminosity to occasional bursting and major flaring, in the single
class of objects we now call magnetars. This is the conclusion we
adopt in this paper. Several authors have written important review
papers on magnetars, their observational properties, and outstanding
questions in the field; \citep[see][]{wt06,mer08,kas10,re11,mer13}.
We note that some alternative models for AXPs and SGRs have been proposed,
including a fall-back disk model that has the sources accreting from
surrounding debris \citep[e.g.][]{eae+07,eeea09}, a massive white
dwarf model \citep[e.g.][]{mrr12}, and also a quark nova model \citep{oln07a,oln07b}.
Although these models are interesting and have their merits, the current
evidence to support these pictures for the overall magnetar population is weak;
however, they may be relevant in describing certain outlier objects. We
consider them no further here but refer the interested reader to the above
references.

With the number of identified magnetars and magnetar candidates having
grown to over two dozen today, the time is ripe for a systematic compilation
of these objects, in the form of the first magnetar catalog, presented
here. Specifically we have collected and compiled a wide variety of information
on the 21 confirmed and 5 unconfirmed magnetars, including their spatial,
spin, and radiative properties across the EM spectrum. Our hope is
that this catalog serves as a useful resource to the magnetar-interested
community, and ultimately helps to identify and highlight important
population properties that could help answer some of the outstanding
questions in magnetar physics. Accompanying this paper is a fully
referenced and linked online version%
\footnote{\url{http://www.physics.mcgill.ca/~pulsar/magnetar/main.html}%
} which is regularly maintained. We note that \citet{mhth05} include
magnetars in their online and published radio pulsar catalog%
\footnote{\url{http://www.atnf.csiro.au/research/pulsar/psrcat/}%
}, however the information compiled there is basic and restricted for the most
part to spatial spin and radio properties.

In $\S$2 we present the catalog in the form of seven data tables
separated by topic. In $\S$3 we provide analysis and discussion of
the magnetar population based on our catalogued data. Finally, concluding
remarks are given in $\S$4.

\section{\label{sec:Data}Data Tables}

\subsection{Table 1: Positions and Proper Motions}

In Table~\ref{tab:Pos} we list the astrometric parameters of the
catalogued magnetars. These include the right ascension and declination
(J2000.0 epoch), the Galactic longitude $l$ and latitude $b$, and
the proper motion $\mu$ in RA and Dec. Measurements of distances
to the magnetars are listed in Table~\ref{tab:Dist}.

The positions listed in this Table are generally those from the literature
with the smallest reported uncertainties. The uncertainties are unchanged
from the original papers and typically, but not necessarily, represent
90\% confidence intervals. In most cases the listed position is from
a \emph{Chandra} observation of the persistent X-ray source, or \emph{Swift}/X-ray
Telescope (XRT) in the case of Swift J1822.3$-$1606. The exceptions
are 4U 0142+61 and SGR 1806$-$20, where the position is of an optical
counterpart, and 1E 1547.0$-$5408, SGR J1745$-$2900, XTE J1810$-$197,
and SGR 1900+14, whose listed positions are of radio counterparts.
Finally, the five candidate magnetars have no confirmed counterparts
at any wavelength, so we list either the best position of the observed
bursts or, in the case of AX J1818.8$-$1559 and AX J1845.0$-$0258,
the \emph{Chandra} position of the unconfirmed persistent X-ray counterpart.

Unlike positions, all of the tabulated proper motion measurements
or upper limits were found in the radio (1E 1547.0$-$5408 and XTE
J1810$-$197) and optical (4U 0142+61, SGR 1806$-$20, 1E 1841$-$045,
SGR 1900+14, and 1E 2259+586) bands (but see \citealt{kch+09} for
proper motion upper limits found in X-ray with \emph{Chandra}). The optical
measurements are all corrected for Galactic rotation, whereas the radio
ones are not. We also caution that the proper motion measurement of
SGR 1900+14 is of its unconfirmed optical counterpart (see Table~\ref{tab:OIR}).

\subsection{Table 2: Timing Properties}

Table~\ref{tab:Time} contains timing parameters for all catalogued
magnetars for which they are available. Specifically, we tabulate
the period $P$ and the epoch at which it was measured, the period
derivative $\dot{P}$ and the range over which it was measured, the
method of measuring $\dot{P}$ (see below), and three physical properties
inferred from $P$ and $\dot{P}$, namely, the surface dipolar magnetic
field strength $B$, defined as $B=3.2\times10^{19}\left(P\dot{P}\right)^{1/2}\, G$;
the spin-down luminosity $\dot{E}$, defined as $\dot{E}=4\pi^{2}I\dot{P}/P^{3}$,
where the moment of inertia $I$ is assumed to be $10^{45}$\,g\,cm$^{2}$;
and the characteristic age $\tau_{c}$, defined as $\tau_{c}=P/2\dot{P}$.
Note that the expression for $B$ assumes simple vacuum dipole radiation
and ignores the potentially important torques due to magnetospheric
variability and the internal superfluid, both of which have been
proposed to be relevant to magnetars \citep{kgw+03,dkg09,akn+13,tlk02,bel09}.

The values of $\dot{P}$ were found using one of two methods. In the
first case (denoted in Table~\ref{tab:Time} by A) $\dot{P}$ is a long-term
average, calculated by fitting a slope to two or more individual
measurements of the period. This was done for sources with only
sparse timing data or, in the cases of 1E 1048.1$-$5937, 1E 1547.0$-$5408,
and SGR 1806$-$20, for sources with large variations in
$\dot{P}$. In the second case (E, ED), $\dot{P}$ was taken from a
phase-coherent timing ephemeris that spans the specified range. If
the ephemeris has higher-order derivatives (denoted by ED), then
the listed value of $\dot{P}$ is only accurate at the period epoch;
otherwise $\dot{P}$ is valid over the entire range. For sources where
multiple phase-coherent timing solutions were found in the literature,
we generally chose the solution from the most recent refereed
publication that covered the most recent glitch-free interval of time,
preferring solutions that covered at least several months. If a publication
presented multiple timing solutions covering the same interval, we
selected the solution that was preferred by the authors. In all cases,
see the references provided for details.

\subsection{Table 3: Quiescent Soft X-ray Properties}

This Table contains the soft X-ray properties of catalog magnetars
in quiescence. To facilitate cross-source comparisons, we generally
report only the phenomenological parameters of an absorbed blackbody
plus power-law model, although in several cases only one of these
two components is required. The columns provided are the neutral hydrogen
column density $N_{\mathrm{H}}$, spectral photon index $\Gamma$,
blackbody temperature $kT$, a second blackbody temperature $kT_{2}$
(only used for CXOU J010043.1$-$721134, for which a blackbody plus
power law was a poor fit to the data), and the absorbed and unabsorbed
fluxes as well as the energy range over which they were derived. We
also include a column for the 2--10\,keV unabsorbed flux, which was
estimated with the WebPIMMS tool%
\footnote{\url{http://heasarc.gsfc.nasa.gov/Tools/w3pimms.html}%
} in cases where the reference gave only absorbed flux or flux in a
different energy range. X-ray luminosities are reported in Table~\ref{tab:Dist}.

The tabulated parameters generally differ in the various papers in
the literature for any given source, so the following explains our
procedure in selecting which properties to catalog. We selected parameters
from publications in which the reported source flux was historically
lowest, in order to ensure as much as possible that the source was
truly in quiescence. In cases where there were multiple publications
with equivalently low flux, we report those model parameters that
had the smallest uncertainties unless more recent observations appeared
more reliable, e.g. were able to better disentangle potentially contaminating
supernova remnant emission. For a majority of the sources this resulted
in the use of spectral parameters obtained from \emph{XMM-Newton}
data, although in several cases the results are taken from \emph{Chandra}
(1E 1048.1$-$5937, Swift J1834.9$-$0846, AX J1845.0$-$0258, and
SGRs 0526$-$66, 1627$-$41, and J1745$-$2900) or archival \emph{ROSAT}
(SGR 0501+4516, XTE J1810$-$197, and Swift J1822.3$-$1606) data
instead. The only other exception is SGR 1806$-$20 for which we use
a model fit derived from simultaneous \emph{Suzaku} and \emph{XMM}
observations. We caution that in general the stated uncertainties,
statistical in nature, may be smaller than the systematic uncertainties
due to calibration and cross-calibration issues; for this reason reported
parameters may not be optimal when considering data from a different
telescope even in the absence of source variability.

There are a few caveats we must make with regards to the flux values
listed in Table~\ref{tab:Xray}. First, although we do list the lowest
reported flux for PSR J1622$-$4950, it is not clear whether the source
had reached quiescence during that observation or whether it was still
fading. Hence, the value we report may be an over-estimation of its
true quiescent flux. Also, note the upper limit for the 2--10\,keV flux of
Swift J1822.3$-$1606 even though it was detected in quiescence.
The reasons for this are that \citet{snl+12} reported the lower bound
for the 0.1--2.4\,keV flux to be zero (likely due to rounding since they do
not claim their result is consistent with a non-detection) and that varying the
spectral parameters within their reported uncertainties changed the
estimated 2--10\,keV flux by over an order of magnitude. We therefore
decided to report the highest such estimated flux as an upper limit.
Additionally, \citet{rie+12} reported somewhat different spectral parameters
for the same observation that gave a 2--10\,keV flux an order of
magnitude greater than the one in the Table; it is this more conservative
value that we use as an upper limit in calculations (including for the
luminosity in Table~\ref{tab:Dist}) and Figures later in this paper. Finally, for
the candidate magnetars AX J1818.8$-$1559 and AX J1845.0$-$0258, we
provide separate spectral parameters and fluxes for single power-law and
single blackbody models, but these results are for \emph{unconfirmed}
quiescent X-ray counterparts that may not be correctly identified.

The parameters in Table~\ref{tab:Xray} are identical to what is
provided in the main Table of our online catalog. However, online
we also provide a table of alternative values including model parameter
results from other observations (e.g. from different telescopes) which
may also be of interest.

\subsection{Table 4: Optical and Near-Infrared Counterparts}

In Table~\ref{tab:OIR} we summarize measurements of catalog magnetars
made in the optical and near-infrared bands. Because magnetars are
typically variable sources at these wavelengths, we list the range
of magnitudes over which they have been detected in the $K_{s}$, $H$,
$J$, $I$, $R$, $V$, $B$, and $U$ bands. We also provide the
limiting magnitudes (usually $3\sigma$ upper limits, but occasionally
2 or $5\sigma$) in cases where observations failed to detect them.

As this Table provides merely a range of values, we reference only
the detections with the lowest and highest reported magnitudes and/or
the non-detection with the highest reported limiting magnitude. In
cases where the same observation was analyzed in both non-refereed
and refereed publications, we considered only the latter for inclusion.
Finally, we must caution that any `non-standard' filter (that is,
any filter other than the eight listed above, such as $K$, $K'$, $z'$,
an \emph{HST} filter, etc.) was assumed to be identical to whichever
standard filter it most closely approximated, with no effort made
to properly convert the magnitude. Therefore, please check the original
references or the online catalog to confirm the filter used.

Seven magnetars have confirmed counterparts in the optical or near-infrared:
4U 0142+61, SGR 0501+4516, 1E 1048.1$-$5937, 1E 1547.0$-$5408, SGR
1806$-$20, XTE J1810$-$197, and 1E 2259+586. Of these, optical pulsations
have been detected from 4U 0142+61, 1E 1048.1$-$5937, and SGR 0501+4516,
of which the latter also shows good evidence for pulsations in the
near-infrared band. There are also suggested counterparts for CXOU
J010043.1$-$721134, 1E 1841$-$045, and SGR 1900+14, but these are
unconfirmed. There was a near-infrared counterpart proposed for 1RXS
J170849.0$-$400910, but \citet{trm+08} disputed the association
when they found multiple fainter sources within the error circle of
its X-ray position. To denote this ambiguity, we report the detected
magnitude of the originally proposed candidate \citep[Star 3 in][]{trm+08}
as an upper limit marked with an asterisk. Similarly, $K_{s}$-band observations
of SGR 1627$-$41 reveal multiple sources that may be the counterpart,
so we list as an upper limit the detected magnitude of the brightest
one \citep[Source C in][]{dcc+09}.

For more information, the online version of this catalog contains
a more comprehensive table of optical and near-infrared counterparts.
It tabulates individual observations of each magnetar, listing the
date of observation, the detected (or limiting) magnitude, and any
non-standard filters that were used.

\subsection{Table 5: Radio and Mid-Infrared Observations}

Table~\ref{tab:RadIR} contains information regarding radio and mid-infrared
observations of catalogued magnetars. For radio observations, we list
all radio frequency ranges in which detections of pulsations have
been reported, as well as the reported dispersion measure (DM). We also
list, where available, the range of detected flux densities in the 1.4 and 2.0
GHz bands; for sources that have never been detected at these wavelengths
we provide an upper limit. Note that transient radio counterparts of SGRs 1806$-$20
and 1900+14 were detected following giant flares \citep{ccr+05,fkb99},
but since no pulsations were ever detected they are not included in
this Table.

For mid-infrared observations, we list the reported fluxes or flux
upper limits for catalogued magnetars at three wavelengths: $4.5\,\mu\mathrm{m}$,
$8.0\,\mu\mathrm{m}$, and $24\,\mu\mathrm{m}$. Note that we are
only concerned with the flux of the point source, so phenomena such
as the infrared ring seen around SGR 1900+14 \citep{wrd+08} are not
included.

\subsection{Table 6: Hard X-ray and Gamma-Ray Observations}

This Table contains the spectral properties of catalog magnetars in
the hard ($>$10\,keV) X-ray and gamma-ray range. The persistent
hard X-ray emission from magnetars can typically be characterized
by a power law, so we report the photon index $\Gamma$ and the unabsorbed
20--150\,keV flux (estimated using WebPIMMS if flux was given for
a different energy range) for both the \emph{pulsed} and \emph{total}
emission, as denoted, respectively, by superscripts $p$ and $t$.
Additionally, because the hard X-ray spectrum is expected to break
or turn over at some point we also list the cut-off energy $E_{\mathrm{cut}}$,
although except for the case of 4U 0142+61, only lower limits are
available.

Most of the hard X-ray data in this Table comes from the \emph{INTEGRAL}
and \emph{Suzaku} telescopes, and we generally tried to include results
from both instruments (in that order) for each source where available.
For results from \emph{INTEGRAL}, we preferred the parameters derived
using the longest integration time, though if it was clear that the
parameters differed between two different time spans we included both
results. As well, in cases where one publication gave multiple parameters
for the same \emph{Suzaku} observation, we chose the one preferred
by the authors. Apart from those two telescopes, \emph{RXTE} data
was used for the pulsed emission from some sources, and the results
for SGR J1745$-$2900 were found with \emph{NuSTAR}. Italicized values
in the Table, seen for SGR 0501+4516, 1E 1547.0$-$5408, and SGR J1745$-$2900,
were taken when the source was in outburst, and here multiple values
of the photon index and flux represent the source fading back into
quiescence. Finally, we must clarify that the inconsistency seen for
1E 2259+586, where the pulsed flux is three times higher than the
upper limit for the total flux, is due to pulsed emission only being
seen by \emph{RXTE} up to $\sim$25\,keV, meaning the extrapolated
flux value reported in the Table must be greatly over-estimated.

Unlike at lower energies, no magnetars have yet been detected in gamma
rays. We therefore provide only upper limits on their 0.1--10\,Gev
flux, taken from Table~1 of \citet{aaa+10}.

\subsection{Table 7: Associations and Distances}

In Table~\ref{tab:Dist} we tabulate distances to catalogued magnetars
and related information. In particular, for each source we list any
objects (e.g. supernova remnants, star clusters, etc.) that are proposed
as associated with it, the age of the supernova remnant (where applicable
and available), the distance measurement, and specifically to which object
the distance is measured (be it the magnetar itself or an associated object).
Associations whose validity has been disputed are noted. We also tabulate
two parameters calculated using the distance $d$: the height above the
Galactic plane $z$, defined as $z=d\sin\left(b\right)$ where $b$ is the
Galactic latitude (see Table~\ref{tab:Pos}); and the quiescent 2--10\,keV
X-ray luminosity $L_{\mathrm{X}}$, defined as $L_{\mathrm{X}}=4\pi
d^{2}F_{\mathrm{X}}$ where $F_{\mathrm{X}}$ is the unabsorbed
2--10\,keV flux (see Table~\ref{tab:Xray}). For sources with no distance
measurements, these derived parameters were estimated assuming a
distance of 10\,kpc. Additionally, since CXOU J010043.1$-$721134 and SGR
0526$-$66 are extragalactic magnetars located in the Magellanic Clouds,
we do not calculate $z$ for them.

In cases where multiple distances to the same source exist in the
literature, we chose the most recently measured value. Usually this
distance was either consistent with earlier measurements or generally
accepted over them among the literature, but for 1E 1048.1$-$5937
and 1E 2259+586 there is some disagreement in the literature between
multiple incompatible distance measurements. For more details, see
the table of alternate values in our online catalog which lists these
other distance measurements with references, or see the discussion
in the papers cited in Table~\ref{tab:Dist}.

\section{\label{sec:Discussion}Discussion}

\begin{figure}
\epsscale{1.15}
\plotone{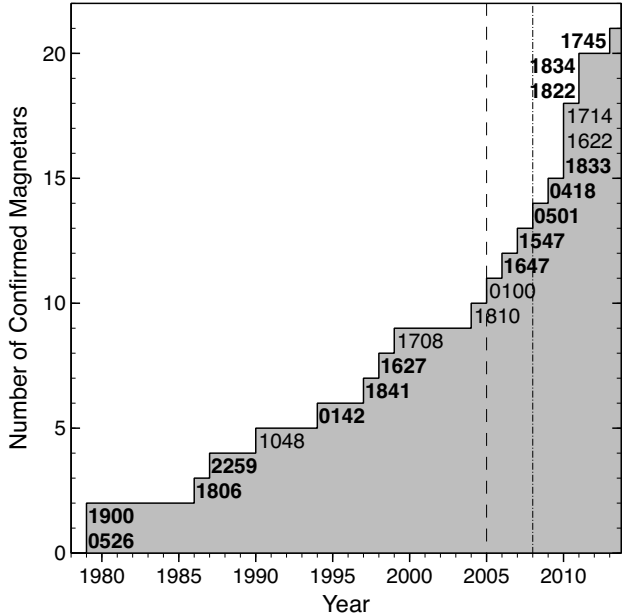}

\caption{\label{fig:MagDisc}Number of confirmed magnetars discovered over
time. Labels in boldface indicate the source was either discovered
or later detected by an all-sky X-ray/soft gamma-ray burst monitor.
The dashed and dot-dashed lines mark the launches of \emph{Swift}
in 2005 and \emph{Fermi} in 2008, respectively.}
\end{figure}

Figure~\ref{fig:MagDisc} shows the accumulated number of known confirmed
magnetars as a function of year up to the present day. The vertical
dashed line shows the launch date of \emph{Swift} with its Burst Alert
Telescope (BAT) onboard \citep{bbc+05}, and the dot-dashed line shows
the launch date of the \emph{Fermi} mission and its Gamma-ray Burst
Monitor \citep[GBM;][]{mlb+09}. It is no coincidence that the slope
of the accumulation increases significantly when BAT became active
and again when GBM turned on, since they are extremely well designed
to detect bright magnetar bursts. In fact they, and previous all-sky
X-ray/soft-gamma ray monitors, were designed to detect gamma-ray bursts,
which are one-time bursters of cosmological origin. Hence these monitors
are specifically designed to view the entire sky in an unbiased fashion,
and so are sensitive to Galactic, repeating bursters regardless of
location in the Galaxy. Thus they have yielded a directionally unbiased
sample of magnetars, selected only for their magnetar activity, namely
bursting. In Figure~\ref{fig:MagDisc}, sources which were either
discovered by an all-sky X-ray/soft gamma-ray monitor, or which were
later detected (and therefore could have been discovered) by one,
are highlighted in italics, for this reason.

Many known magnetars have thus been found via their bursting behavior,
which raises an important point regarding how they are named. Because
burst monitors have tended to find them, magnetars have often been
named with the designation ``SGR'' in recent years (see Tables).
We argue strongly that this naming convention requires amendment because
as discussed in this work and extensively elsewhere \citep[e.g.][]{gkw02,kgw+03,wt06,mer08,kas10,mer13,re11}
the distinction between sources designated as ``AXP'' and ``SGRs''
has been largely erased via the discovery of objects which have properties
previously ascribed to both categories. It is today very hard to classify
some sources as one or the other; rather it has become clear that
there is a continuous spectrum of magnetar-type activity which can
even include some high-$B$ rotation-powered pulsars \citep[e.g. PSR J1846$-$0258;][]{ggg+08}.
Sources discovered via bursting seem like an SGR but may later lie
dormant and burstless for decades and seem like an AXP \citep[e.g. SGR 0526$-$66;][]{kkm+03}.
Meanwhile sources discovered in quiescence and showing no bursts,
therefore initially classified as AXPs, may later begin bursting \citep[e.g. 1E 1547.0$-$5408;][]{gg07,ier+10,kgk+10}.
A source's fixed designation can clearly not depend on behavior that
is constantly evolving. We instead propose a naming scheme that
designates magnetars by the acronym `MG,' analogous to `PSR' as used
for pulsars. A list of MG names is provided in Table~\ref{tab:Name} in the
Appendix. Another possibility would be to keep names as
with other X-ray sources, for which the initial prefix is informative
regarding the discovery telescope, as for, e.g.\ XTE J1810$-$197,
discovered by \emph{RXTE}. We suggest these, and other possible
alternatives, be discussed seriously by the community.

\subsection{Spatial Properties}

\begin{figure*}
\epsscale{0.7}
\plotone{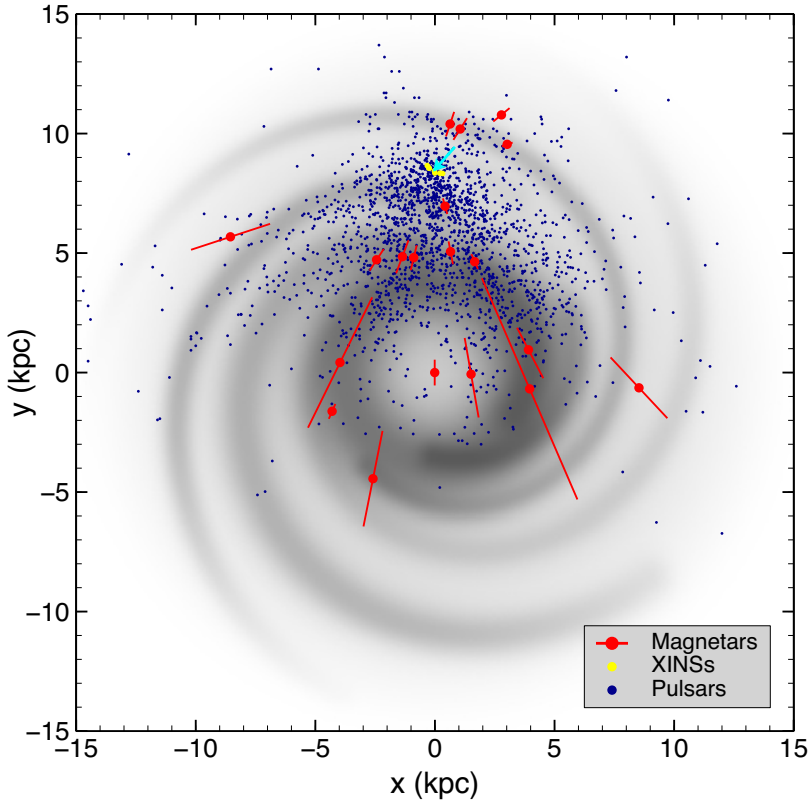}

\caption{\label{fig:MagGalPos}Top-down view of the Galaxy, with the Galactic
Center at coordinates (0,0) and the location of the Sun marked by
a cyan arrow at coordinates (0,8.5). The grayscale shows the distribution
of free electrons given by the model of \citet{cl02}. The magnetars
are denoted by red circles with distance uncertainties indicated by
the lines, the X-ray Isolated Neutron Stars (XINSs) are shown by the
yellow circles near the Sun, and the locations of all other pulsars are
given by the blue dots.}
\end{figure*}

Figure~\ref{fig:MagGalPos} shows a top-down view of the Galactic
Plane with the Galactic Center at coordinate (0,0). The greyscale
is the distribution of free electrons from the model of \citet{cl02}
and delineates the approximate locations of the spiral arms. Galactic
disk radio pulsars from the ATNF catalog%
\footnote{\url{http://www.atnf.csiro.au/research/pulsar/psrcat/}, version 1.47%
} are denoted with blue dots. The so-called `X-ray Isolated Neutron
Stars' (XINSs; see \citealp{krh06,hab07,kap08} for reviews) are shown
in yellow and are without exception very close to the Sun. The magnetars
are shown as red circles, with their estimated distance uncertainties
indicated. Note the magnetar SGR J1745$-$2900 whose location is consistent
with the Galactic Center. This plot clearly indicates the preponderance
of magnetars in the direction of the inner Galaxy, but with several
notable exceptions in the outer Galaxy. The lack of clustering around
the solar system of magnetars, particularly compared with the known
radio pulsar population, suggests that fewer selection effects exist
in the known magnetar population, apart from selection for bursting,
particularly in the \emph{Swift} and \emph{Fermi} eras.

\begin{figure}
\epsscale{1.15}
\plotone{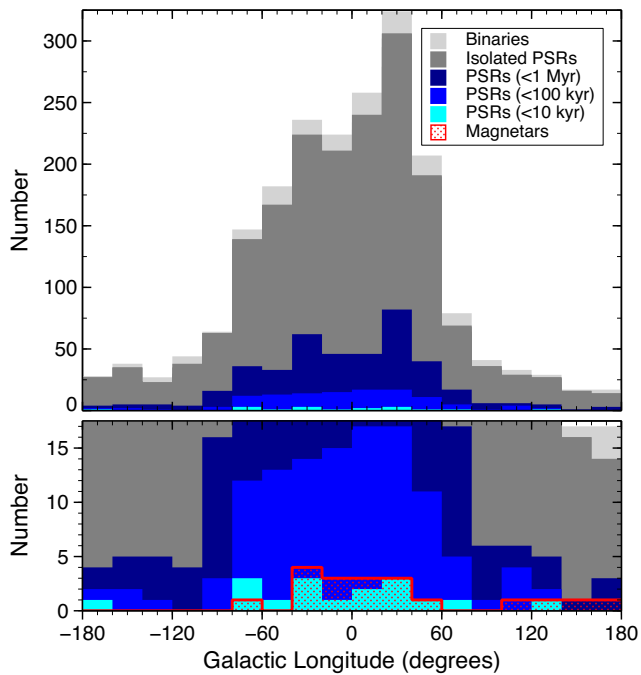}

\caption{\label{fig:HistoLong}Top panel: Distribution in Galactic longitude
$l$ of all Galactic disc pulsars. Young, isolated pulsars are indicated
by the various blue regions ($<$10\,kyr: cyan; $<$100\,kyr: blue;
and $<$1\,Myr: dark blue), with the remaining isolated pulsars and
pulsars in binary systems shown respectively by the grey and light
grey regions. Bottom panel: Zoom-in to better show the distribution
of the magnetars, given by the hatched red region, and the youngest
pulsars.}
\end{figure}

Figure~\ref{fig:HistoLong} presents histograms of the distribution
of ATNF Galactic radio pulsars and magnetars in Galactic longitude
$l$. The radio pulsars are color-coded for age as indicated and the
magnetars are indicated by the hatched red region. As surmised from
Figure~\ref{fig:MagGalPos}, the known Galactic magnetars are more
concentrated in the inner Galaxy, which is not a mere selection effect,
again given the all-sky nature of the burst detectors. While, again,
selection effects in radio pulsar surveys may hinder the detection
of the youngest objects in the very inner Galaxy where multipath scattering
is important \citep{ric90}, we can nevertheless compare the $l$-distributions of
the magnetars and young radio pulsars using a Kolmogorov-Smirnov (KS)
test to see if they are consistent with having been drawn from the
same distribution. For radio pulsars having $\tau<10$\,kyr, we find
a KS probability of the null hypothesis of $p=0.14$, and likewise
we also find $p=0.14$ for $\tau<100$\,kyr. Hence we cannot exclude
that the two distributions are consistent with being drawn from the
same underlying distribution. 

\begin{figure}
\epsscale{1.15}
\plotone{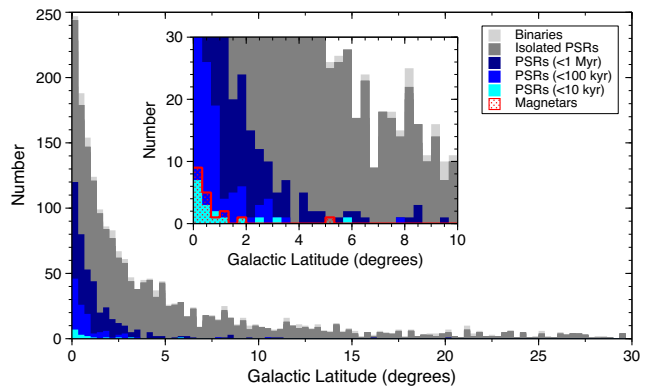}

\caption{\label{fig:HistoLat}Distribution in Galactic latitude $b$ of all
Galactic disc pulsars (colors as in Figure~\ref{fig:HistoLong}).
Inset: Zoom in near the origin with the magnetars shown by the hatched
red region.}
\end{figure}

Figure~\ref{fig:HistoLat} presents histograms of the distribution
of ATNF Galactic disk radio pulsars and magnetars in Galactic latitude
$b$ in degrees, with a zoom-in to the most populated region to better
highlight the magnetars which are relatively few in number. Note that
with the exception of just one magnetar (SGR 0418+5729, but see
$\S$3.2), all known Galactic magnetars lie within 2$\degr$ of the
Galactic Plane, consistent with their interpretation as a population of
young objects. The physical scale height in pc, however, is more relevant
in understanding the Galactic distribution, which we discuss below.

\subsubsection{\label{sec:ScaleHeight}Magnetar Scale Height}

\begin{figure}
\epsscale{1.15}
\plotone{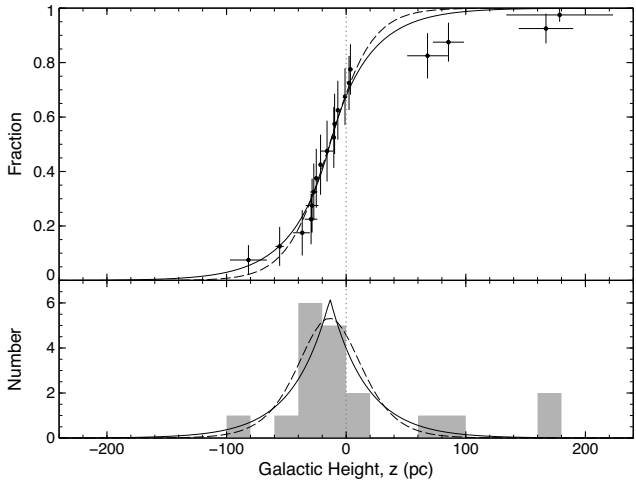}

\caption{\label{fig:GalZ}Top panel: Cumulative distribution function of the
height $z$ above the Galactic plane for the 19 magnetars located
in the Milky Way. Data are fit to an exponential model (solid line)
and a self-gravitating, isothermal disc model (dashed line). See text
for details. Bottom panel: Histogram of the distribution in $z$ of
the Galactic magnetars. Lines are as above.}
\end{figure}

In Figure~\ref{fig:GalZ} (bottom panel) we plot a histogram of the
distribution of magnetars as a function of their height above the
Galactic Plane $z\equiv d\sin(b)$ in pc, where $d$ is the distance
to the object in pc. It is evident that the distribution does not
peak at $z=0$, meaning that simply fitting the distribution to $\exp\left(-\left|z\right|/h\right)$
as is typically done for pulsars will not give an accurate result.
The Sun does not lie in the Galactic plane as defined by the magnetars.
We therefore used two models that included a term for the height of
the Sun: an exponential model and a self-gravitating, isothermal disc
model \citep[e.g.][]{bah84}:
\[
n(z)=n_{0}\exp\!\left(-\frac{\left|z+z_{0}\right|}{h_{e}}\right)\!,\quad\!\!n(z)=n_{0}\,\mathrm{sech}^{2}\!\left(\frac{\left|z+z_{0}\right|}{2h_{s}}\right)\!,
\]
where $h_{e}$ and $h_{s}$ are, respectively, the scale heights for
the exponential and self-gravitating models, and $z_{0}$ is the height
of the Sun above the Galactic plane.

Because of the small number of sources we can work with, as well as
the significant distance uncertainties involved, we constructed and
fit our models to the unbinned cumulative distribution function (top
panel of Figure~\ref{fig:GalZ}) rather than fitting to the histogram
directly. The resulting best-fit values were $h_{e}=30.7\pm5.9$\,pc and
$z_{0}=13.5\pm2.6$\,pc for the exponential model and $h_{s}=17.9\pm3.3$\,pc
and $z_{0}=13.9\pm2.5$\,pc for the self-gravitating model. Note that the
listed $1\sigma$ uncertainties include both the statistical uncertainty
from fitting as well as the $1\sigma$ uncertainty obtained from a
Monte Carlo analysis in which we randomly varied the distance (and
therefore $z$) to each magnetar within their uncertainties. In an effort
to check the stability of our results, we also repeated this procedure for
a few different subsets of the magnetar population. In particular, we tried
fitting the two models to only the 14 Galactic magnetars that have
been detected by all-sky monitors (see Figure~\ref{fig:MagDisc}) since
those sources do not have any sort of directional selection effects.
As well, since the bottom panel of Figure~\ref{fig:GalZ} suggests that
fitting to the cumulative distribution weights the outlying points more
heavily than fitting to the histogram would, we also tested fits excluding
the two sources with $|z|>100$\,pc (SGR 0418+5729 and SGR 1900+14).
We found that these changes tended to decrease $h_{e}$ and $h_{s}$
and increase $z_{0}$; overall the best-fit values for the scale height
varied in the range $\sim$20--31\,pc for $h_{e}$ and $\sim$13--18\,pc
for $h_{s}$, and the best-fit values for the height of the Sun $z_{0}$
ranged from $\sim$13--22\,pc for both models.

For comparison, we repeated the same procedure for all ATNF pulsars
with characteristic age less than 100\,kyr (excluding magnetars)
and found scale heights $h_{e}=61\pm5$\,pc and $h_{s}=39\pm3$\,pc,
approximately twice as large as our results for the magnetars. However,
note that unlike the magnetars, strong selection effects are at work
in shaping the known population of radio pulsars; see, e.g. \citet{fk06}
for a detailed discussion. Indeed, it is generally more difficult
to find faster --- hence typically younger --- radio pulsars closer
to the Galactic Plane because of the deleterious effects of dispersion
smearing and scattering, though recent pulsar surveys of the radio
sky are improving the situation \citep{mlc+01,l++13}. Hence, we may
easily have over-estimated the scale height of young radio pulsars.
Regardless, it is unsurprising that the scale height of magnetars
is smaller or similar to that of young radio pulsars, given that magnetars
are believed to be young neutron stars.

We can also compare our results with measurements in the literature
of the scale height of OB stars, the progenitors of neutron stars.
In particular, \citet{ree00} and \citet{eca06} derived values of
$h_{e}$ ($45\pm20$\,pc and $34\pm2$\,pc, respectively) which
overlap with the upper end of our own range, but other measurements
by \citet{jos07} ($h_{e}=61.4\pm2.6$\,pc) and \citet{mai01} ($h_{s}=34.2$\,pc)
are significantly greater. This discrepancy could argue in favor of
the hypothesis that magnetars are born from massive progenitors \citep{fng+05,mcc+06}
if the OB star scale height depends on stellar mass such that more
massive O stars have a scale height that agrees with that of the magnetars.
Unfortunately, there is no compelling evidence for such a dependence
on stellar mass via spectral type \citep{mas08}, although it cannot
yet be said to be disproven either. Nevertheless, we argue that the
observed magnetar scale height favors massive progenitors. In particular,
9\,$M_{\sun}$ stars have an expected lifetime of about 20\,Myr
\citep{mb81}, so assuming a peculiar velocity of $\sim$5--10\,km\,s$^{-1}$
\citep{gie87} they will have travelled $\sim$70--140\,pc in the
direction perpendicular to the Plane by the end of their lives, significantly
greater than the $\sim$20--30\,pc magnetar scale height. Conversely,
40\,$M_{\sun}$ stars live for approximately 1\,Myr, so given the
same velocity they will travel only $\sim$3--7\,pc during their
life span, a much smaller value that is consistent with the observed
distribution of magnetars.

Finally, we find that our measurement of the height of the Sun above
the Galactic plane $z_{0}$ agrees well with previous measurements,
which generally all fall within the range of 10--30\,pc (e.g. $\sim$10--12\,pc,
\citealp{ree97}; $15\pm3$\,pc, \citealp{cv90}; $16\pm5$\,pc,
\citealp{eca06}; $24.2\pm2.1$\,pc, \citealp{mai01}).

\subsection{\label{sec:Timing}Timing Properties}

\begin{figure}
\epsscale{1.15}
\plotone{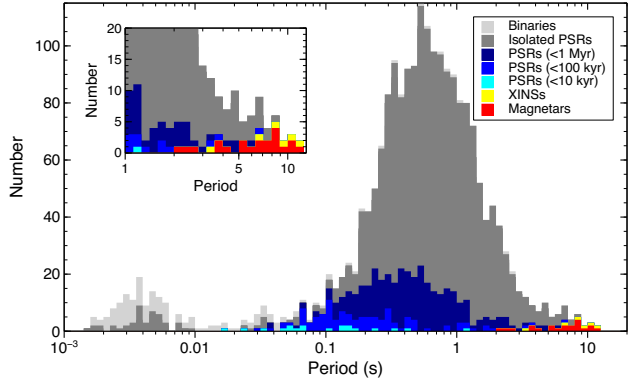}

\caption{\label{fig:HistoP}Histogram showing the distribution in pulse period
of all known radio pulsars (colors as in Figure~\ref{fig:HistoLong}),
XINSs (yellow), and magnetars (red). Inset: Zoom in on $P>1$\,s,
where the magnetars are all located.}
\end{figure}

\begin{figure}
\epsscale{1.15}
\plotone{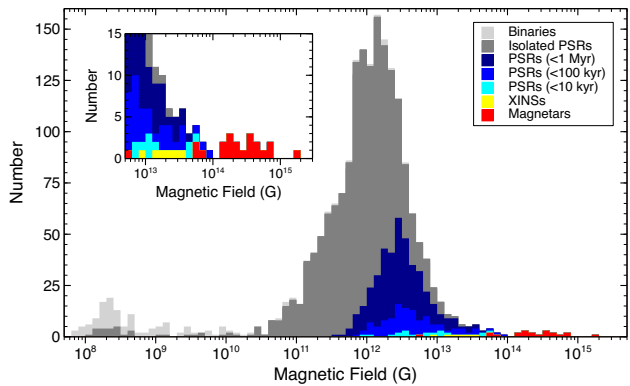}

\caption{\label{fig:HistoB}Histogram showing the distribution in magnetic
field $B$ of all known radio pulsars, XINSs, and magnetars for which
$\dot{P}$ has been measured (colors as in Figure~\ref{fig:HistoP}).
Inset: Zoom in on $B>5\times10^{12}$\,G to better show the distribution
of the magnetars.}
\end{figure}

\begin{figure}
\epsscale{1.15}
\plotone{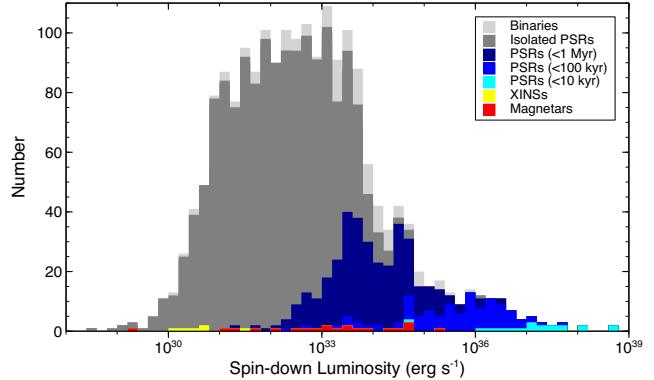}

\caption{\label{fig:HistoEdot}Same as Figure~\ref{fig:HistoB} but for the
spin-down luminosity $\dot{E}$.}
\end{figure}

\begin{figure}
\epsscale{1.15}
\plotone{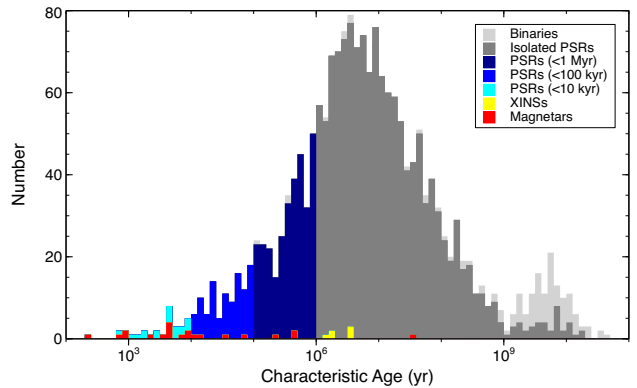}

\caption{\label{fig:HistoAge}Same as Figure~\ref{fig:HistoB} but for the
characteristic age.}
\end{figure}

In Figures~\ref{fig:HistoP}--\ref{fig:HistoAge} we show histograms
of pulse periods and properties inferred from timing for the radio
pulsar population, the XINS, and the magnetars. Figure~\ref{fig:HistoP}
shows the periods, and it is clear that magnetars have longer spin
periods than the vast majority of the radio pulsars, although there
is overlap with the long-period tail of the radio pulsar distribution.
Additionally, the spin periods of the magnetars are very similar to
those of the XINSs. Indeed models of magnetic and thermal evolution
in neutron stars are suggestive of an evolutionary relationship between
magnetars and XINSs, with the latter descendants of the former \citep{vrp+13,ppm+10}.
Notable also is the small range of magnetar periods, especially compared
with those of radio pulsars. The paucity at shorter periods is understood
as being a result of their rapid spin-down due to their high $B$
fields. On the other hand, the reason for the lack of magnetars spin
periods longer than 12\,s is not well established; one possibility
is that by the time objects reach so long a period, their fields have
decayed so much that the hallmark activity and X-ray emission has
ceased \citep[e.g.][]{cgp00}. On the other hand, the longest period
magnetar yet known (1E 1841$-$045) also has the highest persistent
2--10\,keV luminosity (Table~\ref{tab:Xray}). This suggests that
even longer-period magnetars are yet to be found.

In Figure~\ref{fig:HistoB} distributions of the spin-inferred surface
dipolar magnetic field $B$ are shown. Again it is clear that the
typical magnetar field is 2--3 orders of magnitude greater than that
of the typical radio pulsar, and indeed the overlap of the magnetar
field distribution with the high-$B$ tail of the radio pulsar distribution
is relatively small, restricted to three objects (SGR 0418+5729, Swift
J1822.3$-$1606 and 1E 2259+586). Indeed the magnetars largely stand
alone on this plot, with the XINSs having intermediate field values.
Much has been made of the discovery of SGR 0418+5729 \citep{ret+10}
given its low spin-inferred $B$ strength, however Figure~\ref{fig:HistoB}
makes clear that when viewing the overall known magnetar population,
which is largely selected in an unbiased fashion based on burst activity,
low-$B$ objects are the exception.

Interestingly, this Figure also shows that the younger known radio
pulsars tend to have $B$ fields higher than the field of the typical
known radio pulsar. This might naively suggest that radio pulsar magnetic
fields decay with time. On the other hand, higher-field sources spin
down more rapidly, reaching the death line sooner, so the most common
radio pulsar found is likelier to have lower $B$ since it has a longer
lifetime. The small scale height for magnetars described in $\S$\ref{sec:ScaleHeight}
then is consistent with the relative numbers of high-$B$ and low-$B$
magnetars: the objects with the highest fields have the smallest lifetimes
hence have little time to leave their birthplace. Indeed it is unsurprising
that the source with the lowest known $B$ field, SGR J0418+5729,
is also the magnetar furthest from the Galactic Plane (see Table~\ref{tab:Dist}).

Figure~\ref{fig:HistoEdot} shows a histogram of the spin-down luminosity
$\dot{E}$. In this plot, the magnetars are distributed fairly uniformly
but broadly, spanning a full five orders of magnitude. Below we consider
correlations between $\dot{E}$ and radiative properties, but for
the moment we note that the broad range of $\dot{E}$ --- in contrast
to the far narrower and more distinctive range in $B$ --- suggests
that the former does not play a dominant role in the high-energy emission
from magnetars.

Figure~\ref{fig:HistoAge} shows distributions of characteristic
age $\tau_{c}$. As with $\dot{E}$, magnetar ages are uniformly but
broadly distributed. The breadth is interestingly at odds with their
very small Galactic scale height ($\S$\ref{sec:ScaleHeight}), even
given magnetars' relatively low mean velocity \citep{tck13}. This
indicates that the characteristic ages of magnetars are poor proxies
for their true ages. Independent evidence for this is already clear
from the disparity in the characteristic age of 1E 2259+586 (230\,kyr;
see Table~\ref{tab:Time}) compared with the estimated age of
its host supernova remnant CTB 109 (14\,kyr; see Table 7). Note
though that the latter example is extreme; in contrast stands 1E 1841$-$045
whose characteristic age, 4.8\,kyr, is much closer (though still
larger) than the estimated age of its host remnant, Kes 73 (0.5--1\,kyr).
The primary reason for the breadth in characteristic age is unclear.
In some cases it may be at least partially due to fluctuations in
$\dot{P}$ \citep[as in 1E 1048.1$-$5937;][Dib and Kaspi 2013]{gk04}
which could bias a short-term measurement. Alternatively, torque decay
as the magnetic field decays is also a likely factor \citep[e.g.][]{tlk02}.

\begin{figure*}
\epsscale{0.71}
\plotone{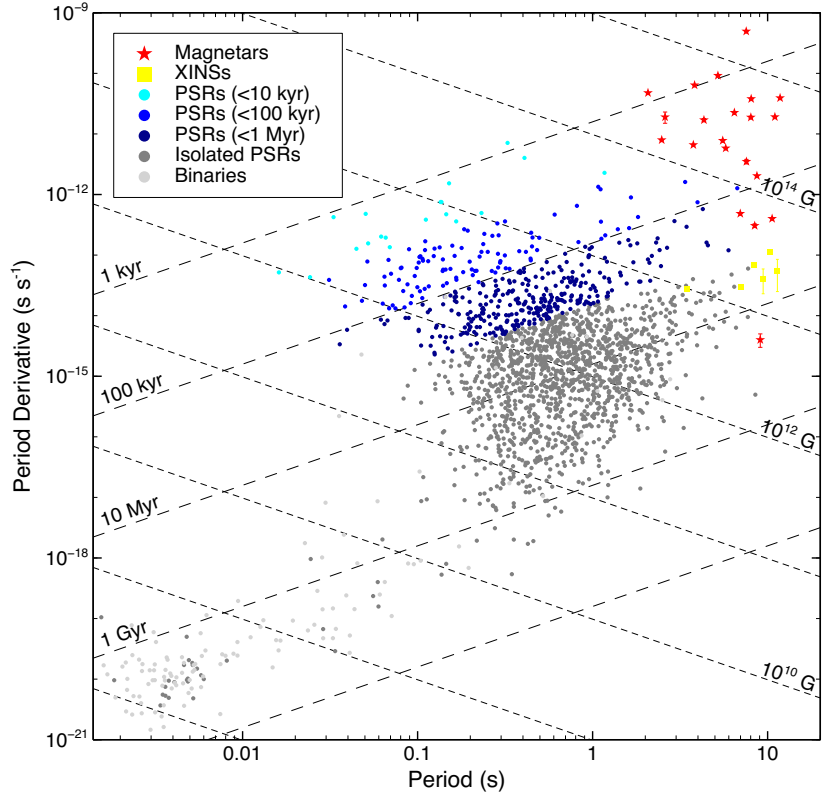}

\caption{\label{fig:PPdot}$P$--$\dot{P}$ diagram for all known radio pulsars
(grey or blue dots as indicated), XINSs (yellow squares), and magnetars
(red stars).}
\end{figure*}

In Figure~\ref{fig:PPdot} we present a $P$--$\dot{P}$ diagram
which includes all catalogued magnetars, XINSs, and radio pulsars
having measured $P$ and $\dot{P}$. This presentation re-emphasizes
the relatively long periods and large spin-down rates of the magnetar
population. Also made clear by this diagram is the overlap in $P$--$\dot{P}$
space between magnetars and some radio pulsars. This is suggestive
of potential magnetar activity from these apparently high-$B$ radio
pulsars. The observed short-lived magnetar activity from rotation-powered
pulsar PSR J1846$-$0258 supports this idea \citep{ggg+08}, as does
apparently enhanced thermal X-ray emission from high-$B$ radio pulsars
compared with that from lower-$B$ radio pulsars of comparable age
\citep{km05,oklk10,zkm+11,ozv+13}. Figure~\ref{fig:PPdot} also
makes clearer that XINS spin properties do not fully overlap with
those of magnetars; the former have smaller spin-down rates hence
smaller inferred $B$. These objects are thus evidence for torque
decay in high-$B$ neutron stars and suggest XINS could be descendants
of magnetars as mentioned above.

\subsection{\label{sec:DiscXray}X-ray Properties}

\begin{figure*}
\plotone{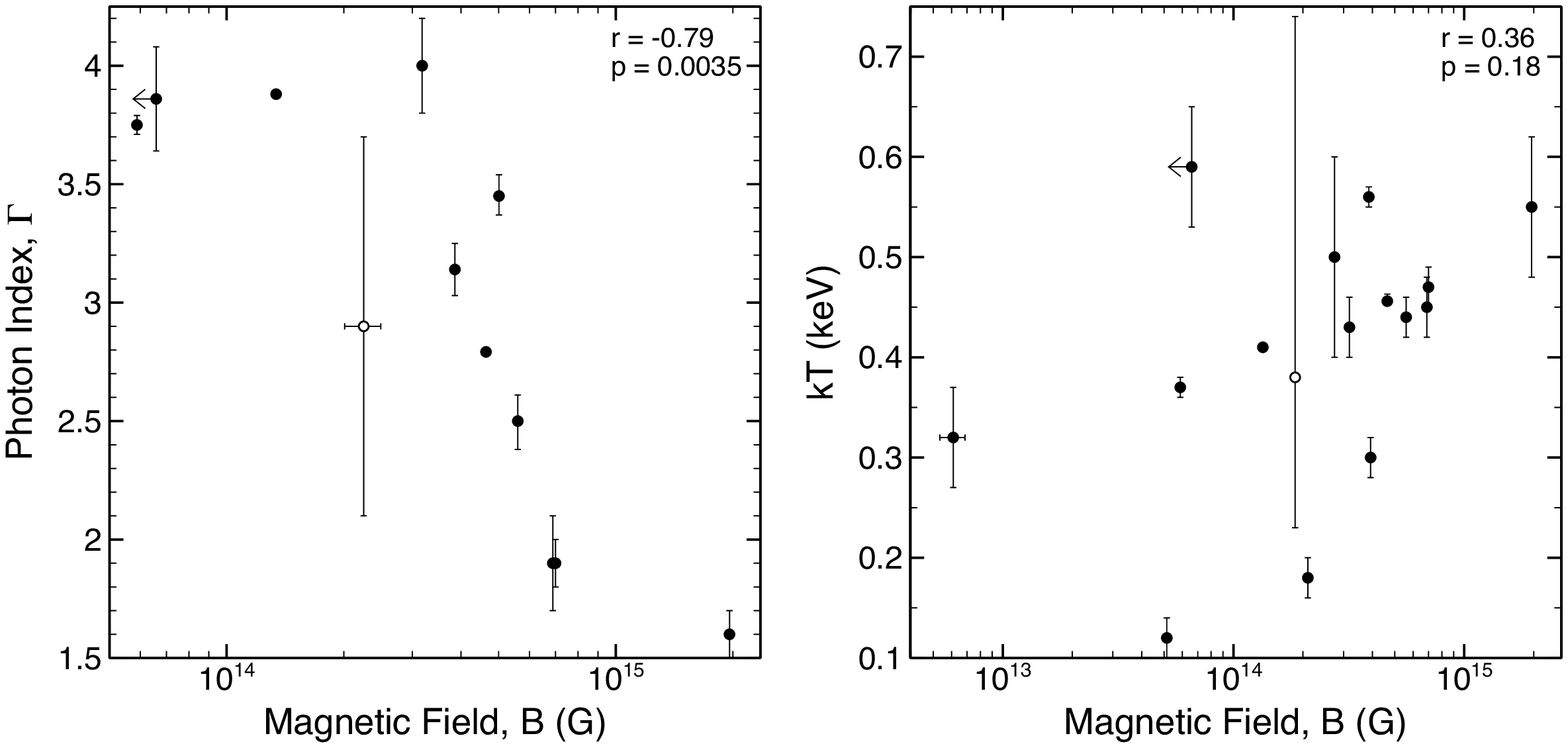}

\caption{\label{fig:GamkTvsB}Photon index $\Gamma$ (left) and blackbody temperature
$kT$ (right) versus magnetic field $B$. The correlation coefficient
$r$ and associated null-hypothesis probability $p$ are shown in
the upper right of each plot. The open circles represent points which
were excluded from the calculation of $r$ due to their large uncertainties
(SGR 1627$-$41 in $\Gamma$ and SGR 0501+4516 in $kT$).}
\end{figure*}

Figure~\ref{fig:GamkTvsB} plots photon index $\Gamma$ and blackbody
temperature $kT$ versus spin-inferred magnetic field $B$ for those
sources which have a power-law or blackbody component in their quiescent
X-ray spectrum (see Table~\ref{tab:Xray}). The left graph shows
evidence of a trend where $\Gamma$ decreases as $B$ increases, previously
identified in \citet{kb10} and in a different but analogous form
by \citet{enm+10b}. Following the example of \citeauthor{kb10} we
attempt to quantify the trend by calculating Pearson's correlation
coefficient, finding $r=-0.79$ (upper limits were included in the
calculation of $r$ by assuming a value of half of the upper limit).
For a sample size of $N=11$, this result gives a (two-tailed) probability for
the null hypothesis of $p=0.0035$, slightly higher than the result
obtained by \citeauthor{kb10} but still near the $3\sigma$ level.
Conversely, examination of the plot on the right for evidence of a correlation
between $kT$ and $B$ revealed none; in particular we obtained $r=0.36$
for $N=15$, giving $p=0.18$ which does not exclude the null hypothesis.
Overall, these results support the ``twisted magnetosphere'' model
of \citet{tlk02}, further developed by \citet{bel09}, which predicts
that a higher $B$ field drives stronger currents in the star's magnetosphere
which in turn produces brighter and harder non-thermal X-ray emission.

\begin{figure*}
\plotone{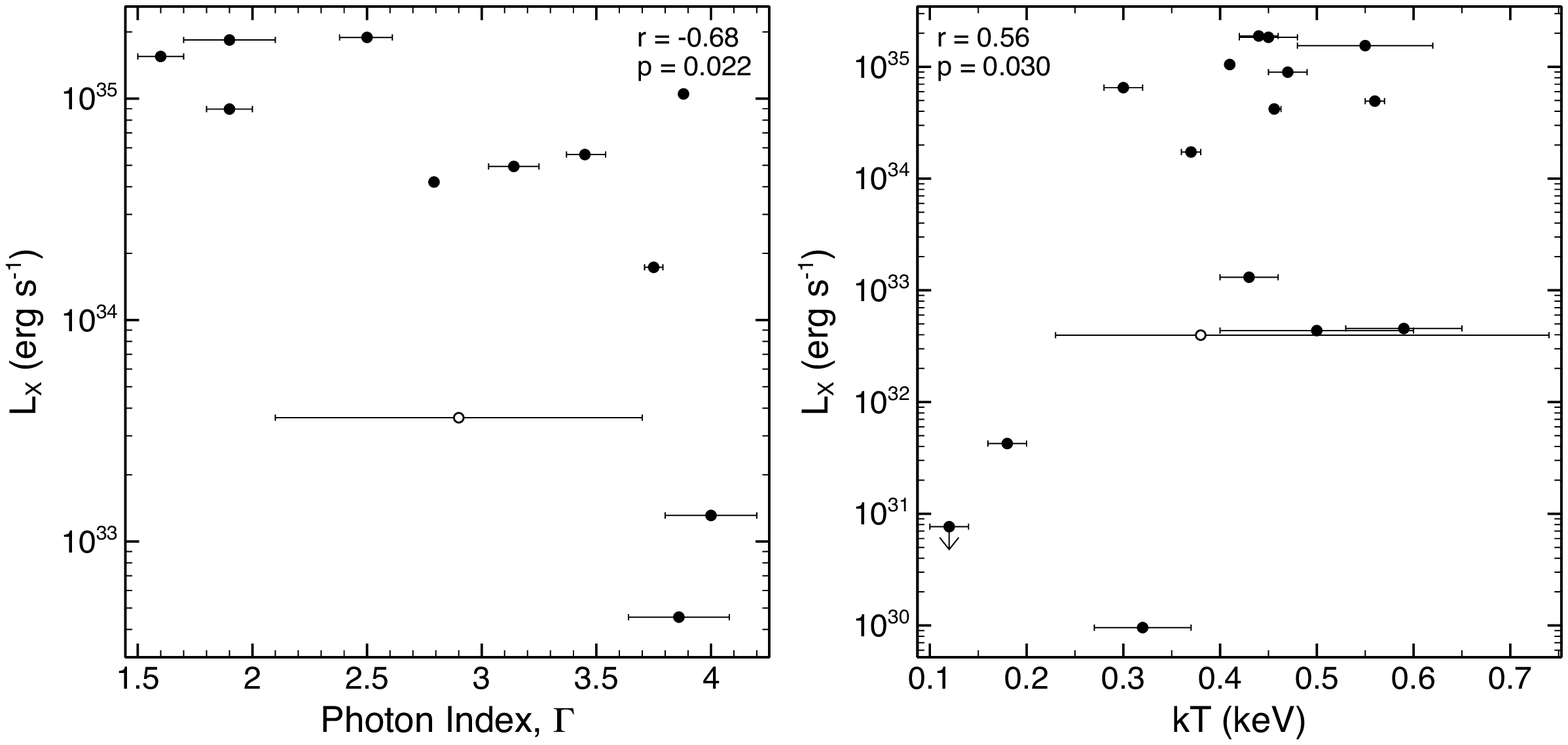}

\caption{\label{fig:LxvsGamkT}Quiescent 2--10\,keV X-ray luminosity $L_{\mathrm{X}}$
vs.\ $\Gamma$ (left) and $kT$ (right). The correlation coefficient
$r$ and null-hypothesis probability $p$ are shown in the upper right
or left of each plot, and the open circles are the same as in Figure~\ref{fig:GamkTvsB}.}
\end{figure*}

In Figure~\ref{fig:LxvsGamkT} we plot $L_{\mathrm{X}}$, the quiescent
X-ray luminosity in the 2--10\,keV energy band, against $\Gamma$
and $kT$ for the same sources as above. We again calculate the correlation
coefficient $r$ but in both cases we derive a null-hypothesis probability
of 0.02--0.03, not low enough to comfortably reject. Certainly a correlation
between $L_{\mathrm{X}}$ and $kT$ is not evident; notice how the
luminosity spans five orders of magnitude at $kT\approx0.3$\,keV.
Likewise, $L_{\mathrm{X}}$ spans more than two orders of magnitude
at $\Gamma\approx3.8$. On the other hand, there does appear to be
an excluded region in the $L_{\mathrm{X}}$ vs.\ $\Gamma$ graph
where one would find lower-luminosity sources with hard power laws
(although given the large uncertainty in $\Gamma$, SGR 1627$-$41
cannot be excluded from encroaching into this region). This cannot
simply be due to a selection effect, because given the same luminosity
a harder source will produce less flux at energies prone to Galactic
absorption than a softer one and should therefore be easier to detect.
As indicated above, a harder spectrum is associated with greater X-ray
luminosity in the twisted magnetosphere model, so such a gap is consistent
with that. However, the model also implies that we should not expect
to see high-luminosity sources with soft power laws. We do note that
a calculation of $r$ excluding the upper-rightmost point (4U 0142+61)
drops the probability of the null hypothesis below 1\% ($r=-0.80$
for $N=10$, $p=0.0054$), although there is no compelling reason
to ignore or discard it.

\begin{figure*}
\epsscale{1.15}
\plotone{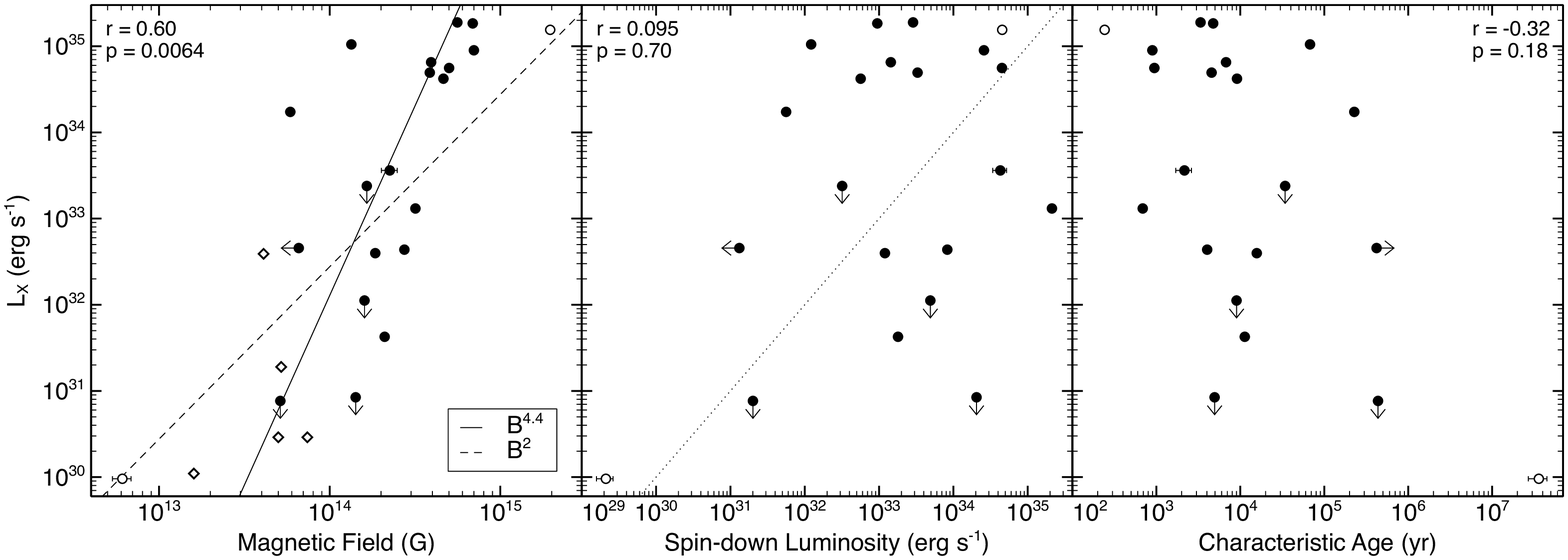}

\caption{\label{fig:LxvsTiming}Left panel: Quiescent 2--10\,keV X-ray luminosity
$L_{\mathrm{X}}$ vs.\ $B$ for the magnetars (solid and open circles)
and select high-$B$ radio pulsars (open diamonds). Data for the radio
pulsars was taken from Table~3 in \citet{akt+12}. The solid and dashed
lines show fits to the data for the relations $L_{\mathrm{X}}\propto B^{4.4}$
and $L_{\mathrm{X}}\propto B^{2}$, respectively. Middle panel:
$L_{\mathrm{X}}$ vs.\ $\dot{E}$. The dotted line marks $L_{\mathrm{X}}/\dot{E}=1$.
Right panel: $L_{\mathrm{X}}$ vs.\ $\tau_{c}$. All panels: The
open circles mark SGRs 0418+5729 and 1806$-$20. Because these two
magnetars lie at opposite corners of each graph, they were excluded
from the calculation of the correlation coefficient $r$, shown together
with the null-hypothesis probability $p$ in the upper left or right
of each plot, to ensure that a correlation did not depend on their
presence.}
\end{figure*}

In the leftmost panel of Figure~\ref{fig:LxvsTiming} we show the
quiescent 2--10\,keV luminosity $L_{\mathrm{X}}$ as a function of
$B$. This plot is an update of Figure~4 from \citet{akt+12}, although
we do not assume the same uncertainties as that paper when drawing
the error bars. The solid and open circles denote the magnetars and
the open diamonds represent the five high-$B$ radio pulsars also
considered by \citeauthor{akt+12}. A possible correlation can be
seen in the data, so as above we investigated it by calculating Pearson's
correlation coefficient and found that it strongly supports the existence
of such a correlation ($r=0.72$ for $N=21$, $p=2.2\times10^{-4}$).
We noticed, however, that there were points in the upper right and
lower left corners of the graph, marked by the open circles, that
could have had a significant impact on the calculation of $r$. Removal
of these extreme points, SGRs 0418+5729 and 1806$-$20, still resulted
in rejection of the null hypothesis ($r=0.60$ for $N=19$, $p=0.0064$).
Furthermore, as in \citeauthor{akt+12} the inclusion of high-$B$
radio pulsars only strengthened the relation ($r=0.73$ for $N=24$,
$p=5\times10^{-5}$), so it appears that there could be a genuine
correlation between $L_{\mathrm{X}}$ and $B$ in high-magnetic-field
neutron stars. There are two other magnetars, 4U 0142+61 and 1E
2259+586, that stand out in the plot with unusually high luminosities given
their lower magnetic fields. This may suggest that their magnetic
fields have strong nondipolar components, not seen in the spin-inferred
field $B$, that would bring the total field strength in line with
the other magnetars of similar $L_{\mathrm{X}}$. Overall, though,
these results support the idea that there is a continuum in the X-ray
luminosities of high-$B$ radio pulsars and magnetars (see \citealp{akt+12}
for further discussion) as expected on physical grounds based on magnetic
dissipation and expected magnetothermal evolution \citep{td96,pmg09}. 

In the middle panel of Figure~\ref{fig:LxvsTiming} we show a plot
of $L_{\mathrm{X}}$ versus spin-down luminosity $\dot{E}$. The panel
shows little more than a scatter plot, as borne out by the correlation
coefficient ($r=0.38$ for $N=21$, $p=0.087$; $r=0.095$, $p=0.70$
with SGRs 0418+5729 and 1806$-$20 removed). This result is expected
in the magnetar model, since the X-ray emission is not powered by
the rotational energy. The rightmost panel of Figure~\ref{fig:LxvsTiming}
presents $L_{\mathrm{X}}$ vs.\ characteristic age, and like the
previous graph there is no visual sign of a strong trend. Naively
calculating the correlation coefficient, however, does show evidence
for a relation ($r=-0.56$ for $N=21$, $p=0.0078$), but it is carried
entirely by SGRs 0418+5729 and 1806$-$20 ($r=-0.32$, $p=0.18$ with
those two points removed). Again, this is unsurprising because not
only are the characteristic ages of magnetars not necessarily good
measures of their true ages as discussed above, but the 2--10\,keV
luminosity is dominated by the non-thermal emission so we do not expect
to detect a cooling trend anyway.

\begin{figure}
\epsscale{1.15}
\plotone{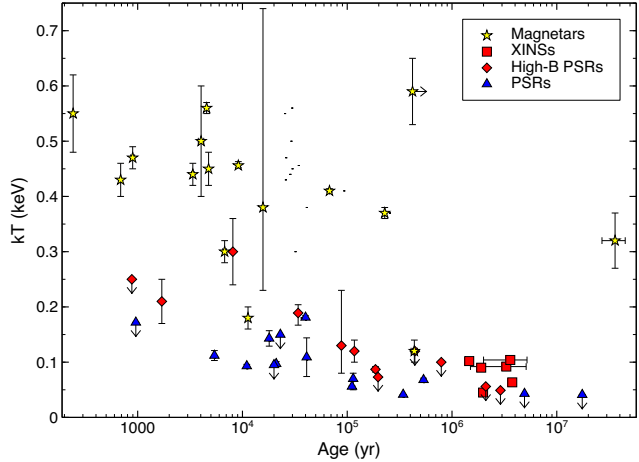}

\caption{\label{fig:kTvsAge}Blackbody temperatures vs.\ characteristic ages
for magnetars (yellow stars), XINSs (red squares), high-$B$ pulsars
($B\geq10^{13}$\,G; red diamonds), and normal pulsars (blue triangles).
Data for the magnetars are taken from this work; for data on the other
sources, see \citet{ozv+13} and references therein, particularly
Table~4 in \citet{zkm+11}.}
\end{figure}

Figure~\ref{fig:kTvsAge} shows a plot of $kT$ versus characteristic
age for magnetars, XINSs, and select radio pulsars, with high-$B$
($\geq10^{13}$\,G) sources shown in red and yellow (with the exception
that the low-field magnetar, SGR 0418+5729, is also shown in yellow).
This is an updated version of Figure~5 in \citet{ozv+13} using the
magnetar data from this work; data for other sources remains unchanged.
As such, our observations and conclusions remain largely unchanged
from the aforementioned paper: there is a general trend for higher-$B$
sources --- of course the magnetars, but also high-$B$ radio pulsars
--- to display greater blackbody temperatures than low-$B$ pulsars
of similar age, suggesting that the magnetic field plays a role in
the observed thermal properties of pulsars. For a more detailed discussion,
see \citet{ozv+13}. Finally, note that SGR 0418+5729, despite having
$B<10^{13}$\,G, is set apart from the other low-$B$ sources by
its much greater $kT$.

\subsection{Multiwavelength Properties}

\begin{figure*}
\epsscale{0.71}
\plotone{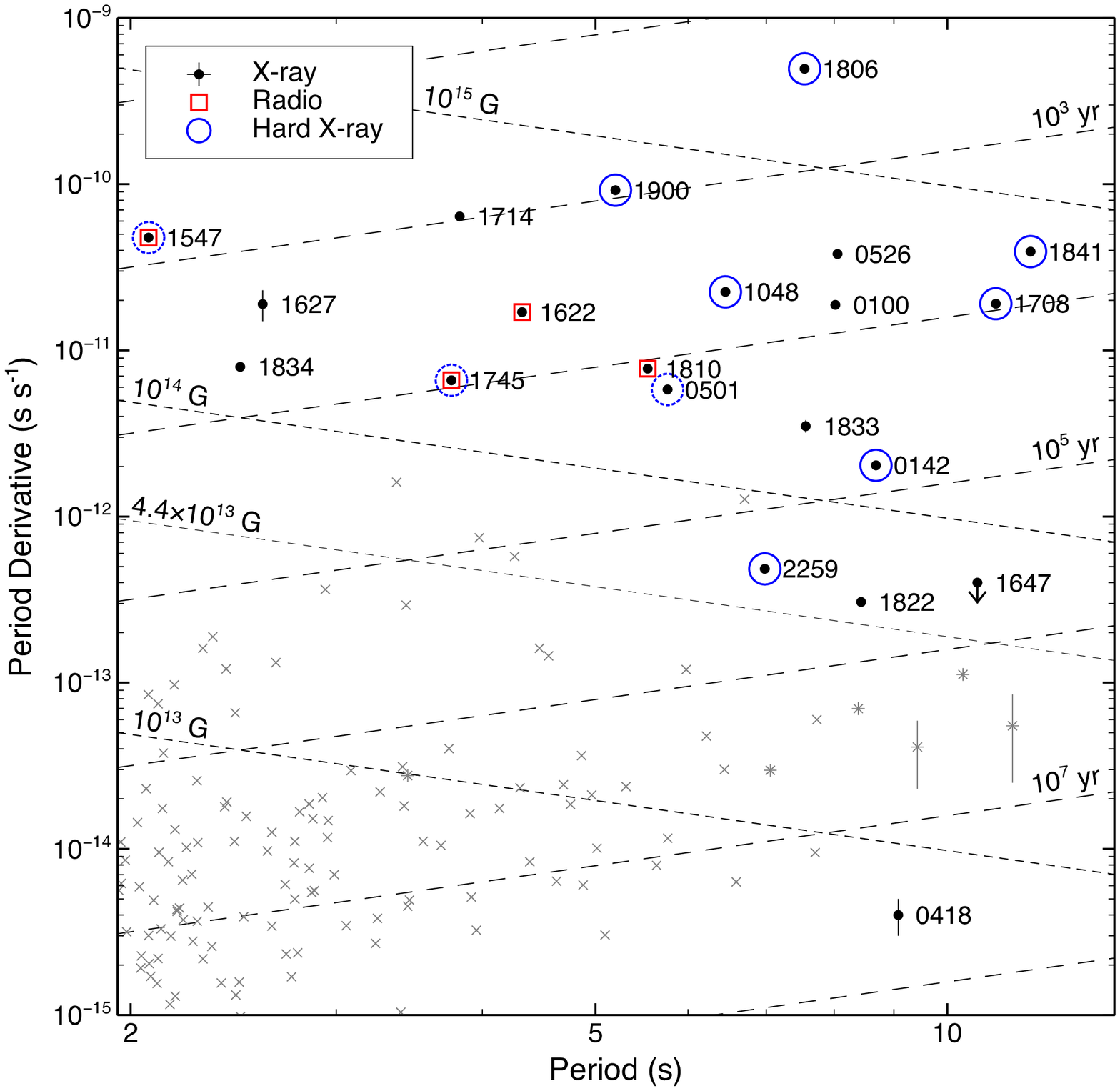}

\caption{\label{fig:MultiPPdot}$P$--$\dot{P}$ diagram showing radio pulsars
(crosses), XINSs (asterisks), and magnetars (circles). Radio-detected
magnetars are marked with red squares. Blue circles denote magnetars
that have been detected in the hard X-ray band ($>$10\,keV), with
a dotted circle indicating that it has been so detected only in outburst.}
\end{figure*}

\begin{figure*}
\epsscale{0.71}
\plotone{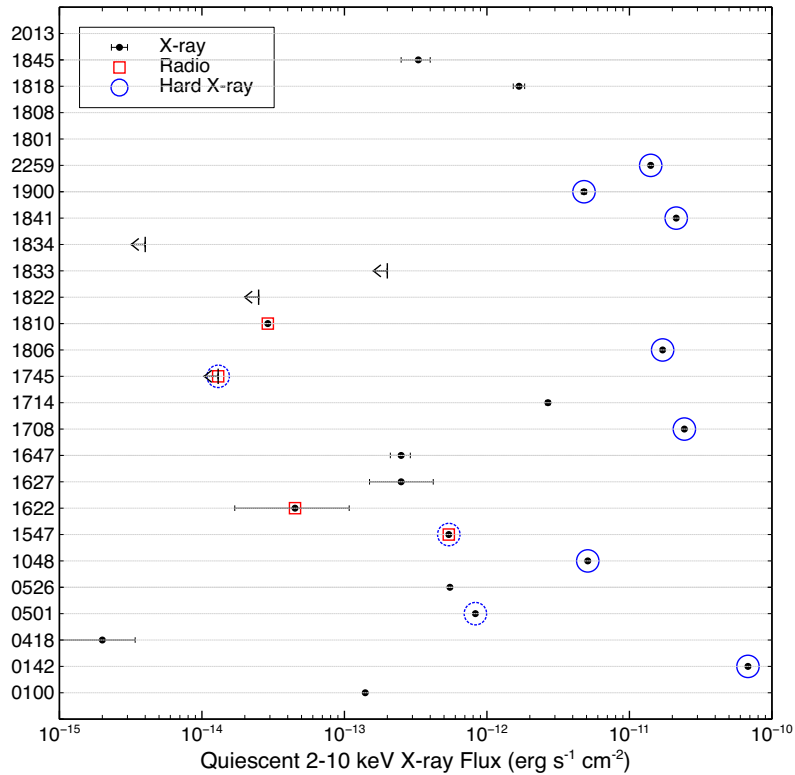}

\caption{\label{fig:MultiFlux}Magnetar detections as a function of quiescent
2--10\,keV X-ray flux $F_{\mathrm{X}}$. Radio and hard X-ray detections
are marked by red squares and blue circles, as described for Figure~\ref{fig:MultiPPdot}.}
\end{figure*}

Figure~\ref{fig:MultiPPdot} shows a $P$--$\dot{P}$ diagram with
radio pulsars, XINSs, and the magnetars indicated, as well as their
detection status in soft X-rays, hard X-rays, and the radio band.
From the plot it is clear that sources detected persistently in hard
X-rays tend to be those with the highest $B$ fields ($10^{14.5}$--$10^{15}$\,G
and above) unless they are particularly distant, e.g. in the Magellanic
Clouds. 4U 0142+61 and 1E 2259+586 are detected in hard X-rays but
have somewhat lower $B$ fields; this further emphasizes their apparently
outlier nature (noted above). Alternatively, from Figure~\ref{fig:MultiFlux},
where the multiwavelength detections of the catalogued magnetars are
shown as a function of quiescent 2--10\,keV X-ray flux $F_{\mathrm{X}}$,
it is clear that only the sources with the highest $F_{\mathrm{X}}$
are detected persistently in hard X-rays. Moreover, hard X-rays are
generally detected in sources in outburst, i.e. when the soft-X-ray
flux is anomalously high. These facts suggest that all magnetars
produce hard X-rays but that current hard X-ray missions do not have
the sensitivity to detect them. NASA's \emph{NuSTAR} mission \citep{hcc+13},
the first focusing hard X-ray telescope, may help in this regard.

The radio emission observed from magnetars is strikingly different
from the hard X-ray behavior. As is clear from Figure~\ref{fig:MultiFlux},
radio emission has only been seen in sources with low $F_{\mathrm{X}}$
when in outburst in spite of extensive radio observations and stringent
upper limits (see Table~\ref{tab:RadIR}) of most of the objects
catalogued including those with the largest values of $F_{\mathrm{X}}$
\citep{bri+06,chk07,lkc+12,tyl13,akn+13}.%
\footnote{We note claimed radio detections of 4U 0142+61, 1E 2259+586 and XTE
J1810$-$197 \citep{mtl12} however the detections have not yet been
confirmed using another observatory.%
} Although beaming may play a role \citep[see discussion in][]{lkc+12},
with increasing statistics, the segregation of radio detections in the
$P$--$\dot{P}$ diagram (Figure~\ref{fig:MultiPPdot}) is interesting.

\citet{rptt12} have suggested that there exists a `fundamental plane'
in magnetar spin and radiative phase space which distinguishes sources
of different radio emission properties. Specifically, they argue on
physical grounds that magnetars with high $\dot{E}$ and low $L_{\mathrm{X}}$
should be radio bright, while low $\dot{E}$, high $L_{\mathrm{X}}$
should not be radio detected. This is in principle an explanation
for the striking asymmetry in the $P$--$\dot{P}$ distribution we
see for radio-emitting sources. On the other hand, the recent non-detection
of magnetar Swift J1834.9$-$0846, which has $\dot{E}/L_{\mathrm{X}}$
where \citet{rptt12} would predict radio emission, argues against
this picture \citep{tyl13,etr+13}. Moreover, the `fundamental plane' picture
also predicts radio emission from the high-$B$ rotation-powered pulsar
PSR J1846$-$0258, which in fact has been shown to be radio quiet
\citep{aklm08}. \citet{rptt12} argue that a previously reported very large
distance (21 kpc) to the source \citep{bh84} together with its presence in
supernova remnant Kes 75 could somehow hinder a radio detection, perhaps
because of a high dispersion measure (DM). However, up-to-date distance
estimates for this pulsar \citep{lt08b,scy+09} place it significantly closer
(5.1--10.9 kpc), and the system's hydrogen column density as measured
with X-ray observations, $N_{\mathrm{H}}=2\textrm{--}4\times10^{22}$\,cm$^{-2}$
predicts, on the basis of an empirical DM vs $N_{\mathrm{H}}$ law
\citep{hnk13}, $DM\simeq600\textrm{--}1300$\,pc\,cm$^{-3}$. This
is well within the range of observed DMs for radio pulsars, particularly
those with only moderately fast rotation periods like the 0.326-s
period of PSR J1846$-$0258. Hence we disagree with the conclusion
of \citet{rptt12} that a radio detection of PSR J1846$-$0258 is
difficult due to its environment. On the other hand, unfortunate radio
beaming, as well as the episodic nature of radio emission from
magnetars, may play a role for this pulsar and Swift J1834.9$-$0846.
Continued radio observations of magnetars in outburst to increase
statistics for radio emission in the population will be helpful for
deciding whether $\dot{E}/L_{\mathrm{X}}$ plays a meaningful role
in radio detectability of magnetars.

\section{Conclusions}

We have compiled the first catalog of all currently known magnetars,
including 21 confirmed sources and 5 candidates. Where available from
the literature we have provided spatial properties (coordinates, proper
motion, distance, and proposed associations), timing data (period,
period derivative, and derived parameters), spectral parameters for
the quiescent soft X-ray emission, and observed properties or upper
limits in the radio, infrared, optical, hard X-ray and gamma-ray bands.
We note that the known magnetar population is relatively free from selection
for location in the Galaxy thanks to the all-sky X-ray monitors that have found
so many of these objects in recent years. We constructed histograms
in Galactic longitude and latitude, spin period $P$, spin-inferred
magnetic field $B$, spin-down luminosity $\dot{E}$, and characteristic
age $\tau_{c}$, to compare the magnetar distributions with the distributions
of the known pulsar population. We measure the scale height of magnetars
for the first time and find it to be smaller than that of OB stars,
supporting the hypothesis that the most massive O stars are magnetar
progenitors. We note the relatively narrow ranges in $P$ and $B$
observed for the magnetars, which stand in contrast to the far wider
ranges in $\dot{E}$ and $\tau_{c}$. That the characteristic age
range is so broad in spite of so small a scale height for these objects
argues that the former is generally a poor age indicator. We confirm
correlations between $\Gamma$ and $B$, previously identified by
\citet{kb10} and \citet{enm+10b}, and $L_{\mathrm{X}}$ and $B$,
previously noted by \citet{akt+12}, and observe an excluded region
in the plot of $L_{\mathrm{X}}$ versus $\Gamma$. Finally, we find
that detections of magnetars in the hard X-ray seem to be correlated
with soft X-ray flux and $B$, while radio detections show, if anything,
the opposite trend. A regularly maintained online version of the catalog
has been made available, with one main table focused on the timing
and X-ray data, and two additional tables for alternative values and
detailed records of optical and near-infrared observations. We plan
to maintain this with regular updates as new magnetar results appear
and encourage the community to provide feedback and suggestions for
improvement on this constantly evolving initiative.

\acknowledgements{We thank J. Lazio for assistance in producing
Figure~\ref{fig:MagGalPos}, as well as N. Murray, L. Drissen, J. Ma{\'\i}z
Apell\'aniz, C. Thompson, B. Gaensler, J. Halpern, S. Tendulkar, R. Duncan,
and the McGill Pulsar Group for helpful discussions. We also thank
K. Boydstun and C. Tam for their work on early versions of the online
magnetar catalog, as well as the magnetar community for their past and
continuing input. V.M.K. holds the Lorne Trottier Chair in Astrophysics and
Cosmology, and a Canada Research Chair, and acknowledges additional
support from an NSERC Discovery Grant and Accelerator Supplement,
from FQRNT via le Centre de Recherche Astrophysique du Qu\'ebec and
the Canadian Institute for Advanced Research.}

\appendix

\section{Magnetar Names}

In Table~\ref{tab:Name} we list the commonly used names (as well as a
few less-common alternatives) of all catalogued magnetars along with our
proposed new names. Said names follow the convention
\textbf{MG~JHHMM+/}$\mathbf{-}$\textbf{DDMM} (here, `MG' stands for
`magnetar') similar to the naming convention for pulsars, and indeed,
for the sake of comparison, we also list the PSR names used by the
ATNF catalog in Table~\ref{tab:Name}. Note that this naming scheme
is only used for confirmed magnetars, although for completeness we also
list the five magnetar candidates in the Table.

\bibliographystyle{apj}
\bibliography{apj-jour,articles,confproc}

\clearpage
\begin{landscape}
\begin{deluxetable*}{lllrrccc}
\tabletypesize{\footnotesize}
\tablewidth{0pt}
\tablecolumns{8}

\tablecaption{\label{tab:Pos}Magnetar Positions and Proper Motions}
\tablehead{
\colhead{Name} & \colhead{Right Ascension\tablenotemark{a}} &
\colhead{Declination\tablenotemark{a}} & \colhead{$l$} & \colhead{$b$} &
\colhead{$\mu_{\mathrm{RA}}$\tablenotemark{b}} &
\colhead{$\mu_{\mathrm{Dec}}$\tablenotemark{b}} & \colhead{References}  \\
\colhead{} & \colhead{(J2000)} & \colhead{(J2000)} &  \colhead{(\degr)} & \colhead{(\degr)} &
\colhead{(mas yr$^{-1}$)} & \colhead{(mas yr$^{-1}$)} & \colhead{}}

\startdata
CXOU J010043.1$-$721134 & $01\ 00\ 43.14(13)$ & $-72\ 11\ 33.8(6)$ & $301.93$ & $-44.92$ & \nodata & \nodata & 1 \\
4U 0142+61 & $01\ 46\ 22.407(28)$\tablenotemark{c} & $+61\ 45\ 03.19(20)$\tablenotemark{c} & $129.38$ & $-0.43$ & $-5.6(1.3)$ & $2.9(1.3)$ & 2, 3 \\
SGR 0418+5729 & $04\ 18\ 33.867(43)$ & $+57\ 32\ 22.91(35)$ & $147.98$ & $+5.12$ & \nodata & \nodata & 4 \\
SGR 0501+4516 & $05\ 01\ 06.76(1)$ & $+45\ 16\ 33.92(11)$ & $161.55$ & $+1.95$ & \nodata & \nodata & 5 \\
SGR 0526$-$66 & $05\ 26\ 00.89(10)$ & $-66\ 04\ 36.3(6)$ & $276.09$ & $-33.25$ & \nodata & \nodata & 6 \\
1E 1048.1$-$5937 & $10\ 50\ 07.14(8)$ & $-59\ 53\ 21.4(6)$ & $288.26$ & $-0.52$ & \nodata & \nodata & 7 \\
1E 1547.0$-$5408 & $15\ 50\ 54.12386(64)$\tablenotemark{d} & $-54\ 18\ 24.1141(20)$\tablenotemark{d} & $327.24$ & $-0.13$ & $4.8(5)$\tablenotemark{f} & $-7.9(3)$\tablenotemark{f} & 8 \\
PSR J1622$-$4950 & $16\ 22\ 44.89(8)$ & $-49\ 50\ 52.7(8)$ & $333.85$ & $-0.10$ & \nodata & \nodata & 9 \\
SGR 1627$-$41 & $16\ 35\ 51.844(20)$ & $-47\ 35\ 23.31(20)$ & $336.98$ & $-0.11$ & \nodata & \nodata & 10 \\
CXOU J164710.2$-$455216 & $16\ 47\ 10.20(3)$ & $-45\ 52\ 16.90(30)$ & $339.55$ & $-0.43$ & \nodata & \nodata & 11 \\
1RXS J170849.0$-$400910 & $17\ 08\ 46.87(6)$ & $-40\ 08\ 52.44(70)$ & $346.48$ & $+0.04$ & \nodata & \nodata & 12 \\
CXOU J171405.7$-$381031 & $17\ 14\ 05.74(5)$ & $-38\ 10\ 30.9(6)$ & $348.68$ & $+0.37$ & \nodata & \nodata & 13 \\
SGR J1745$-$2900 & $17\ 45\ 40.164(2)$\tablenotemark{d} & $-29\ 00\ 29.818(90)$\tablenotemark{d} & $359.94$ & $-0.05$ & \nodata & \nodata & 14 \\
SGR 1806$-$20 & $18\ 08\ 39.337(4)$\tablenotemark{c} & $-20\ 24\ 39.85(6)$\tablenotemark{c} & $10.00$ & $-0.24$ & $-4.5(1.4)$ & $-6.9(2.0)$ & 15, 16 \\
XTE J1810$-$197 & $18\ 09\ 51.08696(28)$\tablenotemark{d} & $-19\ 43\ 51.9315(40)$\tablenotemark{d} & $10.73$ & $-0.16$ & $-6.60(6)$\tablenotemark{f} & $-11.72(1.03)$\tablenotemark{f} & 17 \\
Swift J1822.3$-$1606 & $18\ 22\ 18.00(12)$ & $-16\ 04\ 26.8(1.8)$ & $15.35$ & $-1.02$ & \nodata & \nodata & 18 \\
SGR 1833$-$0832 & $18\ 33\ 44.37(3)$ & $-08\ 31\ 07.5(4)$ & $23.34$ & $+0.02$ & \nodata & \nodata & 19 \\
Swift J1834.9$-$0846 & $18\ 34\ 52.118(40)$ & $-08\ 45\ 56.02(60)$ & $23.25$ & $-0.34$ & \nodata & \nodata & 20 \\
1E 1841$-$045 & $18\ 41\ 19.343(20)$ & $-04\ 56\ 11.16(30)$ & $27.39$ & $-0.01$ & $<$4 & $<$4 & 10, 21 \\
SGR 1900+14 & $19\ 07\ 14.33(1)$\tablenotemark{d} & $+09\ 19\ 20.1(2)$\tablenotemark{d} & $43.02$ & $+0.77$ & $-2.1(4)$ & $-0.6(5)$ & 22, 16 \\
1E 2259+586 & $23\ 01\ 08.295(77)$ & $+58\ 52\ 44.45(60)$ & $109.09$ & $-1.00$ & $-9.9(1.1)$ & $-3.0(1.1)$ & 23, 3 \\
\tableline
SGR 1801$-$23 & $18\ 00\ 59$\tablenotemark{e} & $-22\ 56\ 48$\tablenotemark{e} & $6.91$ & $+0.07$ & \nodata & \nodata & 24 \\
SGR 1808$-$20 & $18\ 08\ 11.2(29.5)$ & $-20\ 38\ 49(414)$ & $9.74$ & $-0.26$ & \nodata & \nodata & 25 \\
AX J1818.8$-$1559 & $18\ 18\ 51.38(4)$ & $-15\ 59\ 22.62(60)$ & $15.04$ & $-0.25$ & \nodata & \nodata & 26 \\
AX 1845.0$-$0258 & $18\ 44\ 54.68(4)$ & $-02\ 56\ 53.1(6)$ & $29.56$ & $+0.11$ & \nodata & \nodata & 27 \\
SGR 2013+34 & $20\ 13\ 56.9(7.3)$ & $+34\ 19\ 48(90)$ & $72.32$ & $-0.10$ & \nodata & \nodata & 28
\enddata

\tablecomments{In this and all subsequent tables, the unconfirmed, candidate magnetars are separated from the confirmed magnetars by a horizontal line.}
\tablenotetext{a}{Positions are of the X-ray source unless otherwise specified.}
\tablenotetext{b}{Proper motions have been corrected for Galactic rotation unless otherwise specified.}
\tablenotetext{c}{Position of the near-infrared counterpart.}
\tablenotetext{d}{Position of the radio counterpart.}
\tablenotetext{e}{See reference for the size and shape of the error box.}
\tablenotetext{f}{Proper motion in the sky frame.}

\tablerefs{
(1) \citet{lfmp02}; (2) \citet{hvk04}; (3) \citet{tck13}; (4) \citet{vck+10}; (5) \citet{gwk+10};
(6) \citet{kkm+03}; (7) \citet{wc02}; (8) \citet{dcrh12}; (9) \citet{ags+12}; (10) \citet{wpk+04};
(11) \citet{mcc+06}; (12) \citet{icp+03}; (13) \citet{hg10b}; (14) \citet{sj13}; (15) \citet{icm+05};
(16) \citet{tck12}; (17) \citet{hcb+07}; (18) \citet{ATel3493}; (19) \citet{gcl+10}; (20) \citet{kkp+12};
(21) \citet{ten13}; (22) \citet{fkb99}; (23) \citet{htv+01}; (24) \citet{cfg+00}; (25) \citet{GCN2351};
(26) \citet{met+12}; (27) \citet{tkgg06}; (28) \citet{sbb+11}}

\end{deluxetable*}
\clearpage

\begin{deluxetable*}{lccccccccc}
\tabletypesize{\footnotesize}
\tablewidth{0pt}
\tablecolumns{10}

\tablecaption{\label{tab:Time}Magnetar Timing Properties}
\tablehead{
\colhead{Name} & \colhead{$P$} & \colhead{Epoch} & \colhead{$\dot{P}$} &
\colhead{$\dot{P}$ Range} & \colhead{Method\tablenotemark{a}} & \colhead{$B$} &
\colhead{$\dot{E}$} & \colhead{$\tau_\mathrm{c}$} & \colhead{References} \\
\colhead{} & \colhead{(s)} & \colhead{(MJD)} & \colhead{($10^{-11}$\,s s$^{-1}$)} &
\colhead{(MJD)} & \colhead{} & \colhead{($10^{14}$\,G)} &
\colhead{($10^{33}$\,erg s$^{-1}$)} & \colhead{(kyr)} & \colhead{}}

\startdata
CXOU J010043.1$-$721134 & 8.020392(9) & 53032 & 1.88(8) & 52044--53033 & A & \phn3.9\phn\phn & \phn\phn1.4\phn\phn\phn\phn & \phn\phn\phn\phn6.8\phn & 1 \\
4U 0142+61 & 8.68832877(2) & 51704 & 0.20332(7) & 51610--53787 & ED & \phn1.3\phn\phn & \phn\phn0.12\phn\phn\phn & \phn\phn\phn68\phd\phn\phn & 2 \\
SGR 0418+5729 & 9.07838822(5) & 54993 & 0.0004(1) & 54993--56164 & E & \phn0.061 & \phn\phn0.00021 & 36000\phd\phn\phn & 3 \\
SGR 0501+4516 & 5.76209653(3) & 54750 & 0.582(3) & 54700--54940 & ED & \phn1.9\phn\phn & \phn\phn1.2\phn\phn\phn\phn & \phn\phn\phn16\phd\phn\phn & 4 \\
SGR 0526$-$66 & 8.0544(2) & 54414 & 3.8(1) & 52152--54414 & A & \phn5.6\phn\phn & \phn\phn2.9\phn\phn\phn\phn & \phn\phn\phn\phn3.4\phn & 5 \\
1E 1048.1$-$5937 & 6.4578754(25) & 54185.9 & $\sim$2.25 & 50473--54474 & A & \phn3.9\phn\phn & \phn\phn3.3\phn\phn\phn\phn & \phn\phn\phn\phn4.5\phn & 6 \\
1E 1547.0$-$5408 & 2.0721255(1) & 54854 & $\sim$4.77 & 54743--55191 & A & \phn3.2\phn\phn & 210\phd\phn\phn\phn\phn\phn & \phn\phn\phn\phn0.69 & 7 \\
PSR J1622$-$4950 & 4.3261(1) & 55080 & 1.7(1) & 54939--55214 & A & \phn2.7\phn\phn & \phn\phn8.3\phn\phn\phn\phn & \phn\phn\phn\phn4.0\phn & 8 \\
SGR 1627$-$41 & 2.594578(6) & 54734 & 1.9(4) & 54620--54736 & A & \phn2.2\phn\phn & \phn43\phd\phn\phn\phn\phn\phn & \phn\phn\phn\phn2.2\phn & 9, 10 \\
CXOU J164710.2$-$455216 & 10.610644(17) & 53999.1 & $<$0.04 & 53513--55857 & A & $<$0.66\phn\phd & \phn$<$0.013\phn\phn\phd & \phn$>$420\phd\phn\phn\phd & 11 \\
1RXS J170849.0$-$400910 & 11.003027(1) & 53635.7 & 1.91(4) & 53638--54015 & ED & \phn4.6\phn\phn & \phn\phn0.57\phn\phn\phn & \phn\phn\phn\phn9.1\phn & 12 \\
CXOU J171405.7$-$381031 & 3.825352(4) & 55272 & 6.40(5) & 54856--55272 & A & \phn5.0\phn\phn & \phn45\phd\phn\phn\phn\phn\phn & \phn\phn\phn\phn0.95 & 13 \\
SGR J1745$-$2900 & 3.7635537(2) & 56424.6 & 0.661(4) & 56406--56480 & E & \phn1.6\phn\phn & \phn\phn4.9\phn\phn\phn\phn & \phn\phn\phn\phn9.0\phn & 14 \\
SGR 1806$-$20 & 7.547728(17) & 53097.5 & $\sim$49.5 & 52021--53098 & A & 20\phd\phn\phn\phn & \phn45\phd\phn\phn\phn\phn\phn & \phn\phn\phn\phn0.24 & 15 \\
XTE J1810$-$197 & 5.5403537(2) & 54000 & 0.777(3) & 53850--54127 & E & \phn2.1\phn\phn & \phn\phn1.8\phn\phn\phn\phn & \phn\phn\phn11\phd\phn\phn & 16 \\
Swift J1822.3$-$1606 & 8.43771958(6) & 55761 & 0.0306(21) & 55758--55991 & ED & \phn0.51\phn & \phn\phn0.020\phn\phn & \phn\phn440\phd\phn\phn & 17 \\
SGR 1833$-$0832 & 7.5654084(4) & 55274 & 0.35(3) & 55274--55499 & ED & \phn1.6\phn\phn & \phn\phn0.32\phn\phn\phn & \phn\phn\phn34\phd\phn\phn & 18 \\
Swift J1834.9$-$0846 & 2.4823018(1) & 55783 & 0.796(12) & 55782--55812 & E & \phn1.4\phn\phn & \phn21\phd\phn\phn\phn\phn\phn & \phn\phn\phn\phn4.9\phn & 19 \\
1E 1841$-$045 & 11.782898(1) & 53824 & 3.93(1) & 53828--53983 & E & \phn6.9\phn\phn & \phn\phn0.95\phn\phn\phn & \phn\phn\phn\phn4.7\phn & 12 \\
SGR 1900+14 & 5.19987(7) & 53826 & 9.2(4) & 53634--53826 & A & \phn7.0\phn\phn & \phn26\phd\phn\phn\phn\phn\phn & \phn\phn\phn\phn0.90 & 20 \\
1E 2259+586 & 6.978948446(4) & 51995.6 & 0.048430(8) & 50356--52016 & ED & \phn0.59\phn & \phn\phn0.056\phn\phn & \phn\phn230\phd\phn\phn & 21 \\
\tableline
SGR 1801$-$23 & \nodata & \nodata & \nodata & \nodata & \nodata & \nodata & \nodata & \nodata & \nodata \\
SGR 1808$-$20 & \nodata & \nodata & \nodata & \nodata & \nodata & \nodata & \nodata & \nodata & \nodata \\
AX J1818.8$-$1559 & \nodata & \nodata & \nodata & \nodata & \nodata & \nodata & \nodata & \nodata & \nodata \\
AX 1845.0$-$0258 & 6.97127(28) & 49272 & \nodata & \nodata & \nodata & \nodata & \nodata & \nodata & 22 \\
SGR 2013+34 & \nodata & \nodata & \nodata & \nodata & \nodata & \nodata & \nodata & \nodata & \nodata
\enddata

\tablenotetext{a}{Method by which $\dot{P}$ was measured. A: Long-term average,
E: Phase-coherent timing ephemeris. ED: Phase-coherent timing ephemeris with
additional higher derivatives.}
\tablenotetext{b}{Other timing solutions with lower $\dot{P}$ are given in \citet{rie+12}
and \citet{snl+12}.}

\tablerefs{
(1) \citet{mgr+05}; (2) \citet{dkg07b}; (3) \citet{rip+13}; (4) \citet{gwk+10}; (5) \citet{tem+09};
(6) \citet{dkg09}; (7) \citet{dksg12}; (8) \citet{lbb+10}; (9) \citet{etm+09}; (10) \citet{ebp+09};
(11) \citet{akac13}; (12) \citet{dkg08}; (13) \citet{sbni10}; (14) \citet{rep+13}; (15) \citet{nmy+09};
(16) \citet{ccr+07}; (17) \citet{snl+12}; (18) \citet{eit+11}; (19) \citet{kkp+12}; (20) \citet{met+06b};
(21) \citet{gk02}; (22) \citet{tkk+98}}

\end{deluxetable*}
\clearpage

\begin{deluxetable*}{lccccccccc}
\tabletypesize{\scriptsize}
\tablewidth{0pt}
\tablecolumns{10}

\tablecaption{\label{tab:Xray}Soft X-ray Properties of Magnetars in Quiescence}
\tablehead{
\colhead{Name} & \colhead{$N_\mathrm{H}$} & \colhead{$\Gamma$} &
\colhead{$kT$} & \colhead{$kT_2$} & \colhead{Abs. Flux\tablenotemark{a}} &
\colhead{Unabs. Flux\tablenotemark{a}} & \colhead{Energy Range} &
\colhead{References} & \colhead{Unabs. Flux\tablenotemark{a}} \\
\colhead{} & \colhead{($10^{22}$\,cm$^{-2}$)} & \colhead{} & \colhead{(keV)} & \colhead{(keV)} &
\colhead{} & \colhead{} & \colhead{(keV)} & \colhead{} & \colhead{(2--10\,keV)}}

\startdata
CXOU J010043.1$-$721134 & $0.063^{+0.020}_{-0.016}$ & \nodata & 0.30(2) & $0.68^{+0.09}_{-0.07}$ & 0.14 & 0.14 & \phn\phd2--10\phd\phn & 1 & 0.14 \\
4U 0142+61 & 1.00(1) & 3.88(1) & $0.410^{+0.004}_{-0.002}$ & \nodata & 58(1) & \nodata & \phn\phd2--10\phd\phn & 2 & 67.9 \\
SGR 0418+5729 & 0.115(6) & \nodata & 0.32(5) & \nodata & 0.012(1) & \nodata & 0.5--10\phd\phn & 3 & $0.0020^{+0.0014}_{-0.0010}$ \\
SGR 0501+4516 & $0.6^{+0.5}_{-0.3}$ & \nodata & $0.38^{+0.36}_{-0.15}$ & \nodata & 1.4 & \nodata & 0.1--2.4\phn & 4 & 0.83 \\
SGR 0526$-$66 & $0.604^{+0.058}_{-0.059}$ & $2.50^{+0.11}_{-0.12}$ & 0.44(2) & \nodata & $1.01^{+0.08}_{-0.13}$ & $1.58^{+0.13}_{-0.20}$ & 0.5--10\phd\phn & 5 & 0.55 \\
1E 1048.1$-$5937 & 0.97(1) & 3.14(11) & 0.56(1) & \nodata & \nodata & 5.1(1) & \phn\phd2--10\phd\phn & 6 & 5.1(1) \\
1E 1547.0$-$5408 & 3.2(2)\tablenotemark{c} & 4.0(2) & 0.43(3) & \nodata & $0.37^{+0.01}_{-0.03}$ & \nodata & 0.5--10\phd\phn & 7 & 0.54 \\
PSR J1622$-$4950\tablenotemark{b} & $5.4^{+1.6}_{-1.4}$ & \nodata & 0.5(1) & \nodata & $0.030^{+0.008}_{-0.006}$ & $0.11^{+0.09}_{-0.04}$ & 0.3--10\phd\phn & 8 & $0.045^{+0.063}_{-0.028}$ \\
SGR 1627$-$41 & 10(2)\tablenotemark{c} & 2.9(8) & \nodata & \nodata & $0.10^{+0.03}_{-0.02}$ & \nodata & \phn\phd2--10\phd\phn & 9, 10 & $0.25^{+0.17}_{-0.10}$ \\
CXOU J164710.2$-$455216 & 2.39(5)\tablenotemark{b} & 3.86(22) & 0.59(6) & \nodata & \nodata & 0.25(4) & \phn\phd2--10\phd\phn & 11 & 0.25(4) \\
1RXS J170849.0$-$400910 & 1.36(4) & $2.792^{+0.008}_{-0.012}$ & $0.456^{+0.007}_{-0.004}$ & \nodata & \nodata & $87.0^{+0.4}_{-0.2}$ & 0.5--10\phd\phn & 12 & 24.3 \\
CXOU J171405.7$-$381031 & $3.95^{+0.15}_{-0.14}$ & $3.45^{+0.09}_{-0.08}$ & \nodata & \nodata & 1.51(3) & 2.68(9) & \phn\phd2--10\phd\phn & 13 & 2.68(9) \\
SGR J1745$-$2900 & \nodata & \nodata & \nodata & \nodata & \nodata & $<$0.013 & \phn\phd2--10\phd\phn & 14 & $<$0.013 \\
SGR 1806$-$20 & 6.9(4) & 1.6(1) & 0.55(7) & \nodata & \nodata & 18(1) & \phn\phd2--10\phd\phn & 15 & 18(1) \\
XTE J1810$-$197 & 0.63(5)\tablenotemark{c} & \nodata & 0.18(2) & \nodata & 0.75 & \nodata & 0.5--10\phd\phn & 16 & 0.029 \\
Swift J1822.3$-$1606 & 0.453(8)\tablenotemark{c} & \nodata & 0.12(2) & \nodata & $0.09^{+0.20}_{-0.09}$ & \nodata & 0.1--2.4\phn & 17 & $<$0.0013\tablenotemark{d} \\
SGR 1833$-$0832 & \nodata & \nodata & \nodata & \nodata & $<$0.02 & $<$0.2 & \phn\phd2--10\phd\phn & 18 & $<$0.2 \\
Swift J1834.9$-$0846 & \nodata & \nodata & \nodata & \nodata & \nodata & $<$0.004 & \phn\phd2--10\phd\phn & 19 & $<$0.004 \\
1E 1841$-$045 & 2.2(1) & 1.9(2) & 0.45(3) & \nodata & \nodata & $43^{+9}_{-12}$ & 0.5--10\phd\phn & 20 & 21.3 \\
SGR 1900+14 & 2.12(8) & 1.9(1) & 0.47(2) & \nodata & \nodata & 4.8(2) & \phn\phd2--10\phd\phn & 21 & 4.8(2) \\
1E 2259+586 & 1.012(7) & 3.75(4) & 0.37(1) & \nodata & 11.5(2) & 14.1(3) & \phn\phd2--10\phd\phn & 22 & 14.1(3) \\
\tableline
SGR 1801$-$23 & \nodata & \nodata & \nodata & \nodata & \nodata & \nodata & \nodata & \nodata & \nodata \\
SGR 1808$-$20 & \nodata & \nodata & \nodata & \nodata & \nodata & \nodata & \nodata & \nodata & \nodata \\
AX J1818.8$-$1559 & 3.6(5) & 1.17(17) & \nodata & \nodata & 1.37(7) & \nodata & \phn\phd2--10\phd\phn & 23 & $1.68^{+0.16}_{-0.15}$ \\
 & 1.6(3) & \nodata & 1.87(12) & \nodata & 1.26(7) & \nodata & \phn\phd2--10\phd\phn & 23 & 1.37(10) \\
AX 1845.0$-$0258 & $7.8^{+2.3}_{-1.8}$ & $1.0^{+0.5}_{-0.3}$ & \nodata & \nodata & 0.28(2) & $0.33^{+0.07}_{-0.08}$ & \phn\phd2--10\phd\phn & 24 & $0.33^{+0.07}_{-0.08}$ \\
 & $5.6^{+1.6}_{-1.2}$ & \nodata & $2.0^{+0.4}_{-0.3}$ & \nodata & 0.26(2) & $0.40^{+0.10}_{-0.11}$ & \phn\phd2--10\phd\phn & 24 & $0.40^{+0.10}_{-0.11}$ \\
SGR 2013+34 & \nodata & \nodata & \nodata & \nodata & \nodata & \nodata & \nodata & \nodata & \nodata
\enddata

\tablenotetext{a}{Fluxes are listed in units of $10^{-12}$\,erg\,s$^{-1}$\,cm$^{-2}$.}
\tablenotetext{b}{The flux of this source was fading and may not yet have reached
quiescence during the observation used.}
\tablenotetext{c}{$N_\mathrm{H}$ was fixed at the best-fit value when fitting the quiescent spectrum.}
\tablenotetext{d}{Elsewhere in this paper we use the more conservative flux upper limit
of $2.5\times10^{-14}$\,erg\,s$^{-1}$\,cm$^{-2}$ for Swift J1822.3$-$1606, derived from
the quiescent parameters given in \citet{rie+12}.}

\tablerefs{
(1) \citet{tem08}; (2) \citet{rni+07}; (3) \citet{rip+13}; (4) \citet{rit+09} (5) \citet{phs+12};
(6) \citet{tgd+08}; (7) \citet{bis+11}; (8) \citet{ags+12}; (9) \citet{eiz+08}; (10) \citet{akt+12};
(11) \citet{akac13}; (12) \citet{rio+07}; (13) \citet{sbni10}; (14) \citet{mgz+13}; (15) \citet{emt+07b};
(16) \citet{ghbb04}; (17) \citet{snl+12}; (18) \citet{eit+11}; (19) \citet{ykk+12}; (20) \citet{ks10};
(21) \citet{met+06b}; (22) \citet{zkd+08}; (23) \citet{met+12}; (24) \citet{tkgg06}}

\end{deluxetable*}
\clearpage

\begin{deluxetable*}{lccccccccc}
\tabletypesize{\footnotesize}
\tablewidth{0pt}
\tablecolumns{10}

\tablecaption{\label{tab:OIR}Optical and Near-Infrared Counterparts of Magnetars}
\tablehead{
\colhead{Name} & \colhead{$K_{s}$} & \colhead{$H$} & \colhead{$J$} &
\colhead{$I$} & \colhead{$R$} & \colhead{$V$} & \colhead{$B$} & \colhead{$U$} & \colhead{References}}

\startdata
CXOU J010043.1$-$721134\tablenotemark{a} & \nodata & \nodata & \nodata & $>$25.9 & \nodata & 24.2--$>$26.2 & $>$25.6 & $>$24.2 & 1, 2 \\
4U 0142+61 & 19.7--20.8 & 20.5--20.9 & 22.0--22.2 & 23.4--24.0\tablenotemark{c} & 24.9--25.6 & 25.3--26.1 & 27.2--28.1 & $>$25.8 & 3--6 \\
SGR 0418+5729 & $>$19.6 & \nodata & $>$27.4 & $>$25.1 & $>$24 & $>$28.6 & \nodata & \nodata & 7--10 \\
SGR 0501+4516 & 18.6--19.7\tablenotemark{c} & \nodata & \nodata & 23.3--24.4\tablenotemark{c} & $>$23.0 & \nodata & $>$26.9 & $>$24.7 & 11--14 \\
SGR 0526$-$66 & \nodata & \nodata & \nodata & $>$26.7 & \nodata & $>$26.6 & $>$24.7 & $>$25.0 & 15 \\
1E 1048.1$-$5937 & 19.4--21.5 & 20.8--$>$23.3 & 21.7--$>$25.0 & 24.9--26.2\tablenotemark{c} & $>$26.0 & $>$25.5 & $>$27.6 & $>$25.7 & 16--21 \\
1E 1547.0$-$5408\tablenotemark{a} & 18.5--$>$21.7 & \nodata & \nodata & \nodata & \nodata & $>$20.4 & $>$20.7 & $>$20.3 & 22--24 \\
PSR J1622$-$4950 & $>$20.7 & \nodata & \nodata & \nodata & \nodata & \nodata & \nodata & \nodata & 25 \\
SGR 1627$-$41 & $\geq$19.1* & $>$19.5 & $>$21.5 & \nodata & \nodata & \nodata & \nodata & \nodata & 26, 27 \\
CXOU J164710.2$-$455216 & $>$21 & \nodata & \nodata & \nodata & \nodata & \nodata & \nodata & \nodata & 28 \\
1RXS J170849.0$-$400910\tablenotemark{b} & $\geq$18.9* & $\geq$20.0* & $\geq$21.9* & $>$25.1 & $>$26.5 & \nodata & \nodata & \nodata & 29--31 \\
CXOU J171405.7$-$381031 & \nodata & \nodata & \nodata & \nodata & \nodata & \nodata & \nodata & \nodata & \nodata \\
SGR J1745$-$2900 & \nodata & \nodata & \nodata & \nodata & \nodata & \nodata & \nodata & \nodata & \nodata \\
SGR 1806$-$20 & 19.3--21.9 & $>$19.5 & $>$21.2 & \nodata & $>$21.5 & \nodata & \nodata & \nodata & 32--34 \\
XTE J1810$-$197 & 20.8--21.9 & 21.5--22.7 & 22.9--23.9 & $>$24.3 & $>$21.5 & $>$22.5 & \nodata & \nodata & 31, 34--38 \\
Swift J1822.3$-$1606 & $>$17.3 & $>$18.3 & $>$19.3 & $>$22.2 & \nodata & \nodata & \nodata & \nodata & 39 \\
SGR 1833$-$0832 & $>$22.4 & \nodata & \nodata & $>$24.9 & \nodata & $>$21.4 & $>$21.3 & $>$22.3 & 40, 41 \\
Swift J1834.9$-$0846 & $>$19.5 & \nodata & \nodata & $>$21.6 & \nodata & \nodata & \nodata & \nodata & 42, 43 \\
1E 1841$-$045\tablenotemark{a} & 19.6--20.5 & 20.8--$>$21.5 & $>$22.1 & \nodata & \nodata & \nodata & \nodata & \nodata & 31, 44 \\
SGR 1900+14\tablenotemark{a} & 19.2--19.7 & \nodata & \nodata & $>$21 & \nodata & \nodata & \nodata & \nodata & 31, 45 \\
1E 2259+586 & 20.4--21.7 & \nodata & $>$23.8 & $>$25.6 & $>$26.4 & \nodata & \nodata & \nodata & 46--48 \\
\tableline
SGR 1801$-$23 & \nodata & \nodata & \nodata & \nodata & \nodata & \nodata & \nodata & \nodata & \nodata \\
SGR 1808$-$20 & \nodata & \nodata & \nodata & \nodata & \nodata & \nodata & \nodata & \nodata & \nodata \\
AX J1818.8$-$1559 & $>$17 & \nodata & \nodata & \nodata & \nodata & \nodata & \nodata & \nodata & 49 \\
AX 1845.0$-$0258 & \nodata & $>$21 & \nodata & \nodata & \nodata & \nodata & \nodata & \nodata & 50 \\
SGR 2013+34 & $>$18.3 & $>$18.5 & $>$19.3 & $>$20.6 & $>$19 & $>$20.2 & $>$21.8 & $>$21.2 & 51--54
\enddata

\tablecomments{We do not distinguish between the standard filters listed and any other
ones such as $K$, $K'$, $z'$, $r'$, etc. See Table~3 of the online catalog or the original
references for further information.}
\tablenotetext{a}{Counterpart is unconfirmed.}
\tablenotetext{b}{The originally proposed counterpart has been disputed by \citet{trm+08}.}
\tablenotetext{c}{Pulsations have been detected in this waveband.}

\tablerefs{
(1) \citet{dv05b}; (2) \citet{dv08}; 
(3) \citet{hvk04}; (4) \citet{dmh+05}; (5) \citet{mkk+05}; (6) \citet{dv06d}; 
(7) \citet{vck+10}; (8) \citet{eit+10}; (9) \citet{dkp11}; (10) \citet{rip+13}; 
(11) \citet{GCN8126}; (12) \citet{GCN8129}; (13) \citet{GCN8160}; (14) \citet{dml+11}; 
(15) \citet{kkv+01}; 
(16) \citet{ics+02}; (17) \citet{wc02}; (18) \citet{dv05a}; (19) \citet{tgd+08}; 
(20) \citet{wbk+08}; (21) \citet{dml+09}; 
(22) \citet{GCN8325}; (23) \citet{mrt+09}; (24) \citet{ATel1909}; 
(25) \citet{ags+12}; 
(26) \citet{wpk+04}; (27) \citet{dcc+09}; 
(28) \citet{ATel910}; 
(29) \citet{icp+03}; (30) \citet{dv06a}; (31) \citet{trm+08}; 
(32) \citet{kot05}; (33) \citet{icm+05}; (34) \citet{ATel195}; 
(35) \citet{ghbb04}; (36) \citet{irm+04}; (37) \citet{rti+04}; (38) \citet{crp+07}; 
(39) \citet{rie+12}; 
(40) \citet{GCN10540}; (41) \citet{gcl+10}; 
(42) \citet{GCN12272}; (43) \citet{kkp+12}; 
(44) \citet{dur05}; 
(45) \citet{GCN1044}; 
(46) \citet{htv+01}; (47) \citet{kgw+03} (48) \citet{tkvd04}; 
(49) \citet{met+12}; 
(50) \citet{isc+04}; 
(51) \citet{GCN4035}; (52) \citet{GCN4036}; (53) \citet{GCN4038}; (54) \citet{GCN4042}} 

\end{deluxetable*}
\clearpage

\begin{deluxetable*}{lccccc|cccc}
\tabletypesize{\footnotesize}
\tablewidth{0pt}
\tablecolumns{10}

\tablecaption{\label{tab:RadIR}Radio and Mid-Infrared Observations of Magnetars}
\tablehead{
\colhead{} & \multicolumn{5}{c}{Radio} & \multicolumn{4}{c}{Mid-Infrared} \\
\colhead{Name} & \colhead{Detection Frequencies} & \colhead{DM} &
\colhead{$S_{1.4\,\mathrm{GHz}}$} & \colhead{$S_{2.0\,\mathrm{GHz}}$} & \colhead{References} &
\colhead{$F_{4.5\,\mathrm{\mu m}}$} & \colhead{$F_{8.0\,\mathrm{\mu m}}$} &
\colhead{$F_{24\,\mathrm{\mu m}}$} & \colhead{References} \\
\colhead{} & \colhead{(GHz)} & \colhead{(cm$^{-3}$\,pc)} & \colhead{($\mu$Jy)} &
\colhead{($\mu$Jy)} & \colhead{} & \colhead{($\mu$Jy)} & \colhead{($\mu$Jy)} &
\colhead{($\mu$Jy)} & \colhead{}}

\startdata
CXOU J010043.1$-$721134 & \nodata & \nodata & \nodata & \nodata & \nodata & \nodata & \nodata & \nodata & \nodata \\
4U 0142+61 & 0.11 & 27 & $<$46 & $<$4.5 & 1--3 & 32.1(2.0) & 59.8(8.5) & $<$38 & 22, 23 \\
SGR 0418+5729 & \nodata & \nodata & \nodata & \nodata & \nodata & \nodata & \nodata & \nodata & \nodata \\
SGR 0501+4516 & \nodata & \nodata & \nodata & $<$40 & 4 & \nodata & \nodata & \nodata & \nodata \\
SGR 0526$-$66 & \nodata & \nodata & \nodata & \nodata & \nodata & \nodata & \nodata & \nodata & \nodata \\
1E 1048.1$-$5937 & \nodata & \nodata & $<$20 & \nodata & 5 & $<$5.2 & $<$21.8 & $<$39 & 24, 25 \\
1E 1547.0$-$5408 & 1.4--8.6, 18.5, 43, 45 & 830(50) & $<$500 -- 4400\tablenotemark{a} & \nodata & 6, 7 & \nodata & \nodata & \nodata & \nodata \\
PSR J1622$-$4950 & 1.4--9.0, 17, 24 & 820(30) & $<$1200 -- 16500\tablenotemark{a} & \nodata & 8--10 & \nodata & \nodata & \nodata & \nodata \\
SGR 1627$-$41 & \nodata & \nodata & $<$80 & \nodata & 11 & \nodata & \nodata & \nodata & \nodata \\
CXOU J164710.2$-$455216 & \nodata & \nodata & $<$40 & \nodata & 12 & \nodata & \nodata & \nodata & \nodata \\
1RXS J170849.0$-$400910 & \nodata & \nodata & $<$20 & \nodata & 5 & $<$120 & $<$170 & $<$590 & 24 \\
CXOU J171405.7$-$381031 & \nodata & \nodata & \nodata & \nodata & \nodata & \nodata & \nodata & \nodata & \nodata \\
SGR J1745$-$2900 & 1.2--8.9, 14.6--20, 22 & 1778(3) & $\sim$90 & $\sim$200 & 13--16 & \nodata & \nodata & \nodata & \nodata \\
SGR 1806$-$20 & \nodata & \nodata & \nodata & $<$6.9 & 2 & \nodata & \nodata & \nodata & \nodata \\
XTE J1810$-$197 & 0.06, 0.35--19, 42, 88.5, 144 & 178(5) & \phn$<$150 -- 13600\tablenotemark{a} & \nodata & 17, 18, 3 & $<$23 & $<$130 & $<$880 & 24 \\
Swift J1822.3$-$1606 & \nodata & \nodata & \nodata & $<$50 & 19 & \nodata & \nodata & \nodata & \nodata \\
SGR 1833$-$0832 & \nodata & \nodata & $<$90 & \nodata & 20 & \nodata & \nodata & \nodata & \nodata \\
Swift J1834.9$-$0846 & \nodata & \nodata & $<$220 & $<$50 & 21 & \nodata & \nodata & \nodata & \nodata \\
1E 1841$-$045 & \nodata & \nodata & $<$20 & $<$10.2 & 5, 2 & \nodata & \nodata & \nodata & \nodata \\
SGR 1900+14 & \nodata & \nodata & \nodata & $<$7.1 & 2 & \nodata & \nodata & \nodata & \nodata \\
1E 2259+586 & 0.06, 0.11 & 79 & \nodata & $<$10.8 & 2, 3 & 6.3(1.0) & $<$20 & \nodata & 26 \\
\tableline
SGR 1801$-$23 & \nodata & \nodata & \nodata & \nodata & \nodata & \nodata & \nodata & \nodata & \nodata \\
SGR 1808$-$20 & \nodata & \nodata & \nodata & \nodata & \nodata & \nodata & \nodata & \nodata & \nodata \\
AX J1818.8$-$1559 & \nodata & \nodata & \nodata & \nodata & \nodata & \nodata & \nodata & \nodata & \nodata \\
AX 1845.0$-$0258 & \nodata & \nodata & $<$20 & $<$9.2 & 5, 2 & \nodata & \nodata & \nodata & \nodata \\
SGR 2013+34 & \nodata & \nodata & \nodata & $<$9.7 & 2 & \nodata & \nodata & \nodata & \nodata
\enddata

\tablenotetext{a}{Since these sources are not always visible in radio, the flux densities
here range from the lowest reported upper limit for a non-detection to the highest
detected value.}

\tablerefs{
(1) \citet{dkh+07}; (2) \citet{lkc+12}; (3) \citet{mtl12}; (4) \citet{GCN8134}; (5) \citet{chk07};
(6) \citet{crhr07}; (7) \citet{crj+08}; (8) \citet{lbb+10}; (9) \citet{kjlb11}; (10) \citet{ags+12};
(11) \citet{ebp+09}; (12) \citet{ATel903}; (13) \citet{sj13}; (14) \citet{efk+13}; (15) \citet{sle+14};
(16) \citet{ATel5076}; (17) \citet{crh+06}; (18) \citet{crp+07}; (19) \citet{rie+12}; (20) \citet{eit+11};
(21) \citet{etr+13}; (22) \citet{wk08}; (23) \citet{wck08}; (24) \citet{wkh07}; (25) \citet{wbk+08};
(26) \citet{kcww09}}

\end{deluxetable*}
\clearpage

\begin{deluxetable*}{lcccccccc|c}
\tabletypesize{\scriptsize}
\setlength{\tabcolsep}{4pt}
\tablewidth{0pt}
\tablecolumns{10}

\tablecaption{\label{tab:HX}Hard X-ray and Gamma-Ray Observations of Magnetars}
\tablehead{
\colhead{} & \multicolumn{8}{c}{Hard X-ray Spectral Parameters} &
\colhead{Gamma Ray\tablenotemark{a}} \\
\colhead{} & \colhead{} & \multicolumn{2}{c}{Pulsed Emission} &&
\multicolumn{2}{c}{Total Emission} & \colhead{} & \colhead{} & \colhead{} \\
\cline{3-4} \cline{6-7}
\colhead{Name} & \colhead{Telescope\tablenotemark{b}} & \colhead{$\Gamma^p$} &
\colhead{$F^p_{\textrm{20--150\,keV}}$\tablenotemark{c}} &&
\colhead{$\Gamma^t$} & \colhead{$F^t_{\textrm{20--150\,keV}}$\tablenotemark{c}} &
\colhead{$E_\mathrm{cut}$ (keV)} & \colhead{References} &
\colhead{$F_{\textrm{0.1--10\,GeV}}$\tablenotemark{c}}}

\startdata
CXOU J010043.1$-$721134 & \nodata & \nodata & \nodata && \nodata & \nodata & \nodata & \nodata & \nodata \\
4U 0142+61 & R, I & 0.40(15) & 2.68(1.34) && 0.93(6) & 9.09(35) & $279^{+65}_{-41}$ & 1 & $<$0.9 \\
 & S & \nodata & \nodata && $0.89^{+0.11}_{-0.10}$ & $\sim$10.3 &  & 2 &  \\
SGR 0418+5729 & \nodata & \nodata & \nodata && \nodata & \nodata & \nodata & \nodata & $<$0.4 \\
SGR 0501+4516 & I, S & \nodata & \nodata && $\mathit{0.79^{+0.20}_{-0.16}}$ & $<$3.5, $\mathit{8.4^{+2.0}_{-1.5}}$ & $>$100 & 3 & $<$1.9 \\
SGR 0526$-$66 & \nodata & \nodata & \nodata && \nodata & \nodata & \nodata & \nodata & \nodata \\
1E 1048.1$-$5937\tablenotemark{d} & \nodata & \nodata & \nodata && \nodata & \nodata & \nodata & \nodata & $<$5.3 \\
1E 1547.0$-$5408 & R, I & $-$$\left(\mathit{0.37^{+0.28}_{-0.20}}\right.$--$\left.\mathit{1.55^{+0.42}_{-0.26}}\right)$ & $\mathit{4.1(9)}$--$\mathit{7.5^{+0.9}_{-1.0}}$ && $\mathit{0.87(7)}$--$\mathit{1.45(4)}$ & $<$1.5, $\mathit{8.0(2.2)}$--$\mathit{25.2(3.7)}$ &  & 5 & $<$10.0 \\
 & S & \nodata & \nodata && $\mathit{1.54^{+0.06}_{-0.05}}$ & $\mathit{17.4^{+1.4}_{-1.8}}$ & $>$200 & 6 &  \\
PSR J1622$-$4950 & \nodata & \nodata & \nodata && \nodata & \nodata & \nodata & \nodata & \nodata \\
SGR 1627$-$41 & \nodata & \nodata & \nodata && \nodata & \nodata & \nodata & \nodata & $<$20.0 \\
CXOU J164710.2$-$455216 & \nodata & \nodata & \nodata && \nodata & \nodata & \nodata & \nodata & $<$10.0 \\
1RXS J170849.0$-$400910 & R, I & 0.86(16) & 2.60(35) && 1.13(6), 1.46(21) & 5.2(1.0), 6.61(23) & $>$300 & 7, 8 & $<$10.0 \\
CXOU J171405.7$-$381031 & \nodata & \nodata & \nodata && \nodata & \nodata & \nodata & \nodata & \nodata \\
SGR J1745$-$2900 & N & \nodata & \nodata && $\mathit{1.47^{+0.46}_{-0.37}}$ & $\mathit{0.67^{+0.20}_{-0.27}}$ & $>$50 & 9 & \nodata \\
SGR 1806$-$20 & I & \nodata & \nodata && 1.5(3), 1.9(2) & 6.0(9), 11(2) & $>$160 & 10, 11 & $<$0.6 \\
 & S & \nodata & \nodata && 1.2(1)--1.7(1) & $\sim$3.8--9.9 &  & 12 &  \\
XTE J1810$-$197 & \nodata & \nodata & \nodata && \nodata & \nodata & \nodata & \nodata & $<$5.0 \\
Swift J1822.3$-$1606 & \nodata & \nodata & \nodata && \nodata & \nodata & \nodata & \nodata & \nodata \\
SGR 1833$-$0832 & \nodata & \nodata & \nodata && \nodata & \nodata & \nodata & \nodata & \nodata \\
Swift J1834.9$-$0846 & \nodata & \nodata & \nodata && \nodata & \nodata & \nodata & \nodata & \nodata \\
1E 1841$-$045 & I & 0.72(15) & $\sim$4.0 && 1.32(11) & $\sim$6.9 & $>$140 & 13 & $<$3.0 \\
 & S & $1.35^{+0.30}_{-0.25}$ & $\sim$2.7 && $1.62^{+0.21}_{-0.22}$ & $\sim$4.6 &  & 14 &  \\
 & N & 0.99(36) & $\sim$3.0 && 1.33(3) & $\sim$8.0 &  & 15 &  \\
SGR 1900+14 & I & \nodata & \nodata && 3.1(5) & 1.6(4) & $>$100 & 16 & $<$0.4 \\
 & S & \nodata & \nodata && 1.2(5)--1.4(3) & $\sim$1.4--3.2 &  & 12 &  \\
1E 2259+586 & R, S & $-1.02(24)$ & $\sim$5.9\tablenotemark{e} && \nodata & $<$2.0 & \nodata & 13, 12 & $<$1.7 \\
\tableline
SGR 1801$-$23 & \nodata & \nodata & \nodata && \nodata & \nodata & \nodata & \nodata & \nodata \\
SGR 1808$-$20 & \nodata & \nodata & \nodata && \nodata & \nodata & \nodata & \nodata & \nodata \\
AX J1818.8$-$1559 & \nodata & \nodata & \nodata && \nodata & \nodata & \nodata & \nodata & \nodata \\
AX 1845.0$-$0258 & \nodata & \nodata & \nodata && \nodata & \nodata & \nodata & \nodata & \nodata \\
SGR 2013+34 & \nodata & \nodata & \nodata && \nodata & \nodata & \nodata & \nodata & \nodata
\enddata

\tablecomments{Values in italics were measured when the source was in outburst.}
\tablenotetext{a}{Gamma ray flux upper limits are taken from \citet{aaa+10}.}
\tablenotetext{b}{R: \emph{RXTE}, I: \emph{Integral}, S: \emph{Suzaku}, N: \emph{NuSTAR}.}
\tablenotetext{c}{Hard X-ray and gamma ray fluxes are in units of $10^{-11}$\,erg\,s$^{-1}$\,cm$^{-2}$.}
\tablenotetext{d}{1E~1048.1$-$5937 was detected in hard X-rays with \emph{INTEGRAL}
by \citet{lwr08}, but no spectral information was given.}
\tablenotetext{e}{Pulsed emission from 1E~2259+586 was only observed by \emph{RXTE}
up to $\sim$25\,keV, so the extrapolated 20--150\,keV pulsed flux should not be considered
reliable.}

\tablerefs{
(1) \citet{dkh+08}; (2) \citet{emn+11}; (3) \citet{rit+09}; (4) \citet{ern+10}; (5) \citet{khdu12};
(6) \citet{enm+10a}; (7) \citet{gri+07}; (8) \citet{dkh08}; (9) \citet{mgz+13}; (10) \citet{mgmh05};
(11) \citet{mhs+05}; (12) \citet{enm+10b}; (13) \citet{khdc06}; (14) \citet{mks+10}; (15) \citet{ahk+13};
(16) \citet{gmte06}}

\end{deluxetable*}
\clearpage

\begin{deluxetable*}{lcccccccc}
\tabletypesize{\scriptsize}
\tablewidth{0pt}
\tablecolumns{9}

\tablecaption{\label{tab:Dist}Magnetar Associations and Distances}
\tablehead{
\colhead{Name} & \colhead{Proposed Associations} & \colhead{SNR Age} & \colhead{References} & \colhead{Distance} & \colhead{Measured To} & \colhead{Reference} & \colhead{$z$} & \colhead{$L_\mathrm{X}$\tablenotemark{a}} \\
\colhead{} & \colhead{} & \colhead{(kyr)} & \colhead{} & \colhead{(kpc)} & \colhead{} & \colhead{} & \colhead{(pc)} & \colhead{}}

\startdata
CXOU J010043.1$-$721134 & SMC & \nodata & 1 & 62.4(1.6) & SMC & 28 & \nodata & \phn65\phd\phn\phn\phn\phn\phn \\
4U 0142+61 & \nodata & \nodata & \nodata & 3.6(4) & 0142+61 & 29 & $-27(3)$ & 105\phd\phn\phn\phn\phn\phn \\
SGR 0418+5729 & \nodata & \nodata & \nodata & $\sim$2 & Perseus Arm & 30 & $\sim$$180$ & \phn\phn0.00096 \\
SGR 0501+4516 & SNR HB 9\tablenotemark{b} & 4--7 & 2, 3 & $\sim$2 & Perseus Arm & 31 & $\sim$$68$ & \phn\phn0.40\phn\phn\phn \\
SGR 0526$-$66 & LMC, SNR N49\tablenotemark{b}, SL 463 & $\sim$4.8 & 4--6 & 53.6(1.2) & LMC & 32 & \nodata & 189\phd\phn\phn\phn\phn\phn \\
1E 1048.1$-$5937 & GSH 288.3$-$0.5$-$28\tablenotemark{b} & \nodata & 7 & 9.0(1.7) & 1048.1$-$5937 & 29 & $-82(15)$ & \phn49\phd\phn\phn\phn\phn\phn \\
1E 1547.0$-$5408 & SNR G327.24$-$0.13 & \nodata & 8 & 4.5(5) & 1547.0$-$5408 & 33 & $-10.3(1.1)$ & \phn\phn1.3\phn\phn\phn\phn \\
PSR J1622$-$4950 & SNR G333.9+0.0 & $<$6 & 9 & $\sim$9 & J1622$-$4950 & 34 & $\sim$$-16$ & \phn\phn0.44\phn\phn\phn \\
SGR 1627$-$41 & CTB 33, MC $-71$, SNR G337.0$-$0.1 & \nodata & 10, 11 & 11.0(3) & G337.0$-$0.1 & 11 & $-21.4(6)$ & \phn\phn3.6\phn\phn\phn\phn \\
CXOU J164710.2$-$455216 & Westerlund 1 & \nodata & 12 & 3.9(7) & Westerlund 1 & 35 & $-29(5)$ & \phn\phn0.45\phn\phn\phn \\
1RXS J170849.0$-$400910 & \nodata & \nodata & \nodata & 3.8(5) & J170849.0$-$400910 & 29 & $2.4(3)$ & \phn42\phd\phn\phn\phn\phn\phn \\
CXOU J171405.7$-$381031 & SNR CTB 37B & $0.65^{+2.50}_{-0.30}$ & 13, 14 & $\sim$13.2 & CTB 37B & 36 & $\sim$$86$ & \phn56\phd\phn\phn\phn\phn\phn \\
SGR J1745$-$2900 & Galactic Center & \nodata & 15 & $\sim$8.5 & Galactic Center & 37 & $\sim$$-7.0$ & \phn$<$0.11\phn\phn\phn\phd \\
SGR 1806$-$20 & W31, MC 13A, Star cluster & \nodata & 16, 17 & $8.7^{+1.8}_{-1.5}$ & Star cluster & 38 & $-36.7^{+6.3}_{-7.6}$ & 163\phd\phn\phn\phn\phn\phn \\
XTE J1810$-$197 & \nodata & \nodata & \nodata & $3.5^{+0.5}_{-0.4}$ & J1810$-$197 & 39 & $-9.7^{+1.1}_{-1.4}$ & \phn\phn0.043\phn\phn \\
Swift J1822.3$-$1606 & M17 & \nodata & 18 & 1.6(3) & M17 & 18 & $-28.5(5.3)$ & \phn$<$0.0077\phn\phd \\
SGR 1833$-$0832 & \nodata & \nodata & \nodata & \nodata & \nodata & \nodata & $\sim$$3.6$ & \phn$<$2.4\phn\phn\phn\phn\phd \\
Swift J1834.9$-$0846 & SNR W41 & $\sim$100 & 19, 20 & 4.2(3) & W41 & 40 & $-25(2)$ & \phn$<$0.0084\phn\phd \\
1E 1841$-$045 & SNR Kes 73 & 0.5--1 & 21, 22 & $8.5^{+1.3}_{-1.0}$ & Kes 73 & 22 & $-0.97^{+0.11}_{-0.15}$ & 184\phd\phn\phn\phn\phn\phn \\
SGR 1900+14 & Star cluster & \nodata & 23 & 12.5(1.7) & Star cluster & 41 & $167(23)$ & \phn90\phd\phn\phn\phn\phn\phn \\
1E 2259+586 & SNR CTB 109 & 14(2) & 24, 25 & 3.2(2) & CTB 109 & 42 & $-55.6(3.5)$ & \phn17\phd\phn\phn\phn\phn\phn \\
\tableline
SGR 1801$-$23 & \nodata & \nodata & \nodata & \nodata & \nodata & \nodata & $\sim$$12$ & \nodata \\
SGR 1808$-$20 & \nodata & \nodata & \nodata & \nodata & \nodata & \nodata & $\sim$$-45$ & \nodata \\
AX J1818.8$-$1559 & \nodata & \nodata & \nodata & \nodata & \nodata & \nodata & $\sim$$-44$ & \phn20\phd\phn\phn\phn\phn\phn \\
AX 1845.0$-$0258 & SNR G29.6+0.1 & $<$8 & 26 & $\sim$8.5 & Scutum Arm & 43 & $\sim$$16$ & \phn\phn2.9\phn\phn\phn\phn \\
SGR 2013+34 & W58 & \nodata & 27 & $\sim$8.8 & W58 & 27 & $\sim$$-16$ & \nodata
\enddata

\tablenotetext{a}{2--10\,keV X-ray luminosity in units of $10^{33}$\,erg\,s$^{-1}$.
No uncertainties have been included.}
\tablenotetext{b}{The proposed association with this source has been disputed.}

\tablerefs{
(1) \citet{lfmp02}; (2) \citet{GCN8149}; (3) \citet{lt07}; (4) \citet{cdt+82}; (5) \citet{khg+04};
(6) \citet{phs+12}; (7) \citet{gmo+05}; (8) \citet{gg07}; (9) \citet{ags+12}; (10) \citet{wkv+99b};
(11) \citet{ccdd99}; (12) \citet{mcc+06}; (13) \citet{nbi+09}; (14) \citet{hg10a}; (15) \citet{mgz+13};
(16) \citet{fmc+99}; (17) \citet{ce04}; (18) \citet{snl+12}; (19) \citet{tllw07}; (20) \citet{kkp+12};
(21) \citet{vg97}; (22) \citet{tl08}; (23) \citet{vhl+00}; (24) \citet{fg81}; (25) \citet{spgb13};
(26) \citet{ggv99}; (27) \citet{sbb+11}; (28) \citet{hgd12b}; (29) \citet{dv06b}; (30) \citet{vck+10};
(31) \citet{lkb+11}; (32) \citet{hgd12a}; (33) \citet{tve+10}; (34) \citet{lbb+10}; (35) \citet{kd07};
(36) \citet{tl12}; (37) \citet{sj13}; (38) \citet{bcfc08}; (39) \citet{mcr+08}; (40) \citet{lt08a};
(41) \citet{dfk+09}; (42) \citet{kf12}; (43) \citet{tkk+98}}

\end{deluxetable*}
\clearpage
\end{landscape}

\begin{deluxetable}{lccc}
\tabletypesize{\footnotesize}
\tablewidth{0pt}
\tablecolumns{4}
\tablecaption{\label{tab:Name}Magnetar Names}
\tablehead{
\colhead{Current Name} & \colhead{Alternate Current Name} &
\colhead{MG Name} & \colhead{ATNF (PSR) Name}}

\startdata
CXOU J010043.1$-$721134 & \nodata & MG J0100$-$7211 & PSR J0100$-$7211 \\
4U 0142+61 & \nodata & MG J0146+6145 & PSR J0146+6145 \\
SGR 0418+5729 & \nodata & MG J0418+5732 & PSR J0418+5732 \\
SGR 0501+4516 & \nodata & MG J0501+4516 & PSR J0501+4516 \\
SGR 0526$-$66 & \nodata & MG J0526$-$6604 & PSR J0525$-$6607 \\
1E 1048.1$-$5937 & \nodata & MG J1050$-$5953 & PSR J1048$-$5937 \\
1E 1547.0$-$5408 & SGR J1550$-$5418 & MG J1550$-$5418 & PSR J1550$-$5418 \\
PSR J1622$-$4950 & \nodata & MG J1622$-$4950 & PSR J1622$-$4950 \\
SGR 1627$-$41 & \nodata & MG J1635$-$4735 & PSR J1635$-$4735 \\
CXOU J164710.2$-$455216 & \nodata & MG J1647$-$4552 & PSR J1647$-$4552 \\
1RXS J170849.0$-$400910 & \nodata & MG J1708$-$4008 & PSR J1708$-$4009 \\
CXOU J171405.7$-$381031 & \nodata & MG J1714$-$3810 & PSR J1714$-$3810 \\
SGR J1745$-$2900 & SGR J1745$-$29 & MG J1745$-$2900 & PSR J1745$-$2900 \\
SGR 1806$-$20 & \nodata & MG J1808$-$2024 & PSR J1808$-$2024 \\
XTE J1810$-$197 & \nodata & MG J1809$-$1943 & PSR J1809$-$1943 \\
Swift J1822.3$-$1606 & \nodata & MG J1822$-$1604 & PSR J1822$-$1606 \\
SGR 1833$-$0832 & \nodata & MG J1833$-$0831 & PSR J1833$-$0831 \\
Swift J1834.9$-$0846 & \nodata & MG J1834$-$0845 & PSR J1834$-$0845 \\
1E 1841$-$045 & \nodata & MG J1841$-$0456 & PSR J1841$-$0456 \\
SGR 1900+14 & \nodata & MG J1907+0919 & PSR J1907+0919 \\
1E 2259+586 & \nodata & MG J2301+5852 & PSR J2301+5852 \\
\tableline
SGR 1801$-$23 & \nodata & \nodata & \nodata \\
SGR 1808$-$20 & \nodata & \nodata & \nodata \\
AX J1818.8$-$1559 & GRB 071017 & \nodata & \nodata \\
AX 1845.0$-$0258 & \nodata & \nodata & PSR J1845$-$0256 \\
SGR 2013+34 & GRB 050925 & \nodata & \nodata
\enddata

\end{deluxetable}

\end{document}